\documentclass[final]{aa}  

\usepackage{graphicx}
\usepackage{txfonts}
\usepackage{lipsum}
\usepackage{subcaption} 
\usepackage{booktabs}
\usepackage[toc,page]{appendix}
\usepackage{subcaption}
\usepackage{multicol}
\usepackage{multirow}
\usepackage{lscape}             
\usepackage{placeins}           
                                
\begin{document}

\title{Constraints on Cosmic-Rays and the MeV Emission from Jupiter's Atmosphere}

\author{Dimitris Tsatsis\inst{1} \and Thomas Siegert\inst{1} \and Volodymyr Savchenko\inst{2} \and Saurabh Mittal\inst{1} \and Laura Eisenberger\inst{1} \and Mika\,A. Gelowicz\inst{3,1} \and Rudi Reinhardt\inst{1} \and Manja\,L. Zimmerer\inst{1} \and Tristan Bouchet\inst{1} \and Tomohiko Oka\inst{1} \and Manuel\,R.\,H.\,W. Skalka}

\institute{Julius-Maximilians-Universität Würzburg, Fakultät für Physik und Astronomie, Institut für Theoretische Physik und Astrophysik, Lehrstuhl für Astronomie, Emil-Fischer-Str. 31, D-97074 Würzburg, Germany
\and
ISDC, Department of Astronomy, University of Geneva, Chemin d’Ecogia 16, 1290 Versoix, Switzerland
\and
Rheinische Friedrich-Wilhelms-Universität Bonn, Argelander-Institut für Astronomie (AIfA), Auf dem Hügel 71, D-53121 Bonn,
Germany\\
\email{dimitrios.tsatsis@uni-wuerzburg.de}}



\abstract{The Cosmic $\gamma$-ray Background (CGB) in the MeV regime remains among the least understood components of diffuse cosmic radiation, primarily due to instrumental limitations in detecting faint isotropic emission at MeV energies. We present an indirect method for probing the isotropic emission from soft X-rays to $\gamma$-rays through scattered emission in Jupiter’s atmosphere. Using archival observations from SPI (spectrometer aboard INTEGRAL), spanning 22 years of serendipitous Jupiter observations, we analyse the planet as a moving $\gamma$-ray point source in the 30 keV–8 MeV range. Our analysis focuses on Jupiter’s $\gamma$-ray albedo, created as a result of high-energy photon and cosmic-ray interactions. We performed detailed Monte Carlo simulations to model the yield of Jupiter’s atmosphere to incident radiation, including photons, protons, electrons, positrons and alpha particles. We simulate a 1000 km deep spherical atmosphere with 1 km shell resolution, based on Galileo probe measurements. By comparing empirical and astrophysical models to the SPI data, we find no statistically significant emission from Jupiter. We derive 2$\sigma$ upper limits across the 30 keV–8 MeV band, which we compare to cutoff power-law, reflected CGB, cosmic-ray-induced, and long-lived mediator dark matter models. We constrain the bremsstrahlung luminosity from Jupiter in the MeV range to $10^{18}$ erg/s, which is close to inferred in-situ measurements obtained by Juno/JEDI. While the expected emission from the CGB and cosmic-rays is a factor of $10^4$ and 100 below our upper limits respectively, this work establishes planetary atmospheres as promising indirect probes of diffuse $\gamma$-ray backgrounds for future MeV missions.}




\maketitle
\nolinenumbers

\section{Introduction}\label{sec:intro}

The CGB in the hard X-ray to soft $\gamma$-ray regime (hundreds of keV to tens of MeV) remains one of the least understood components of diffuse cosmic radiation \citep{Ajello_2008, Universe, CGB}. The CGB could hold hidden information about a large number of physical processes that could take place in the $\gamma$-ray region, such as red-shifted and broad $\gamma$-ray lines from the population of type Ia supernovae, $\gamma$-ray bursts and $\gamma$-rays from AGN \citep{Ruiz} that could be good candidates for high-energy cosmic rays (CR) and neutrinos \citep{murase2024high}. It  is also an excellent messenger to possibly uncover signs for the dark matter phenomenon responsible for various gravitational effects \citep{DarkM}.

Direct measurements of the MeV CGB are challenging because $\gamma$-ray telescopes in this energy range are dominated by instrumental background and lack the sensitivity to robustly separate faint isotropic signals from instrumental background \citep{siegert2022diffuse}. Early measurements by HEAO-1 \citep[HEAO-1]{HOE-1} established the hard X-ray background up to several hundred keV, while the Solar Maximum Mission (SMM) provided initial, albeit questionable, constraints into the MeV regime through broad-band $\gamma$-ray spectroscopy \citep{SMM}. Later, COMPTEL onboard the Compton Gamma Ray Observatory (CGRO) extended diffuse measurements to energies of several tens of MeV \citep{COMPTEL}, revealing a spectral transition between the hard X-ray and GeV $\gamma$-ray backgrounds but with substantial uncertainties in normalisation and source decomposition. Even more recently, the Spectrometer onboard INTEGRAL \citep[SPI;][]{Vedrenne2003}, improved spectral resolution in the sub-MeV to MeV band and enabled studies of diffuse Galactic emission, including the 511 keV positron-annihilation line \citep{IntegralLeg, yoneda2025imaging}, but robust measurements of the isotropic MeV background remain strongly limited by instrumental sensitivity \citep{churazov2008earth, Tuerler2010}.

To overcome this, we used an indirect approach, using Jupiter as a reflector to concentrate the isotropic CGB into a point source, allowing spectral measurements influenced by Jupiter’s atmospheric composition. The  planet's atmosphere is continuously bombarded by Galactic CRs, producing secondary $\gamma$-ray emission, including a potential contribution around 511 keV. Detecting such emission from Jupiter would provide an observational leverage for both the reflected CGB and CR-induced $\gamma$-ray albedo.

Composed mainly of hydrogen and helium, Jupiter lacks a solid surface, with its atmosphere exhibiting dynamic weather patterns, including powerful storms and lightning. Multiple measurements with different instruments indicate that different emission mechanisms could be expected from Jupiter. With Juno/JEDI \citep{Mauk2017} electron and ion populations in Jupiter's polar regions were detected, suggesting high energy particles up to the MeV range. Consistent with this, Chandra and XMM-Newton observations have revealed persistent, pulsating X-ray auroral hot spots at both of Jupiter's poles, produced by charge-exchange emission from ions with energies reaching tens of MeV \citep{Chandra}. Moreover, Fermi-LAT has been used to search for dark matter annihilation signatures from Jupiter, exploiting its large mass and proximity \citep{JupiterDM}. Each of these phenomena are expected to produce a measurable signature in the MeV range, making it a natural regime in which to look for and disentangle them. Motivated by this, we analyse the $\gamma$-ray emission from Jupiter, using Monte Carlo simulations of Jupiter's atmosphere, to estimate the CGB reflection and CR induced $\gamma$-ray albedo, spanning a wide energy range. Estimates obtained from these simulations allow us to obtain a method of distinction between intrinsic (CR induced albedo) and extrinsic (CGB reflection) emission from Jupiter and in the future from Solar System bodies in general.

Jupiter is not a stationary point source, but it is moving along the ecliptic, which is not standard when it comes to observation and data analysis processes with SPI. We used more than 22 years of INTEGRAL/SPI data of serendipitous observations of Jupiter, during its orbit, when the location of the SPI measurement corresponds with the Jupiter's position as it moves in the sky. These observations arise because SPI's field of view (16 $\times$ 16 degrees), subtends a solid angle large enough that pointings dedicated to other targets have a non-negligible probability of containing Jupiter at the time of observation. As Jupiter's position changes along the ecliptic it is observed by SPI in different positions across the 22 years, through observation of other targets. This approach differs from the standard planet-centric technique, in which the source is placed at a fixed position relative to the instrument. This allows consistent imaging to be applied across the dataset. In our case, since the pointings were not designed to observe Jupiter, its offset is different for every single pointing, set purely by the chance alignment between Jupiter's position and INTEGRAL's pointing direction at that time. We treat every pointing as an independent measurement before combining them into the final result.

This paper is structured as follows. In Sect.\,\ref{sec:data_analysis}, we describe the observational dataset and the analysis procedure used to extract the $\gamma$-ray emission from Jupiter with SPI, including the treatment of Jupiter as a moving source and the spectral fitting framework. In Sect.\,\ref{DM}, we introduce the intrinsic $\gamma$-ray emission from Jupiter. In Sect.\,\ref{sec:simulations}, we present Monte Carlo simulations of Jupiter’s atmospheric response to incident photons and cosmic rays. The observational and simulation constraints we find on the reflected CGB emission, cosmic-ray-induced $\gamma$-rays, and dark matter scenarios are presented in Sect.\,\ref{sec:results}. Following that, in Sect.\,\ref{Discussion}, we provide an interpretation of our results. Finally, Sect.\,\ref{Conclusions} offers a summary of our main conclusions and outlines future directions for this work.

\section{Data analysis}\label{sec:data_analysis}
\subsection{Obtaining the dataset}
\begin{figure}[h!]
    \centering
    \includegraphics[width=\linewidth]{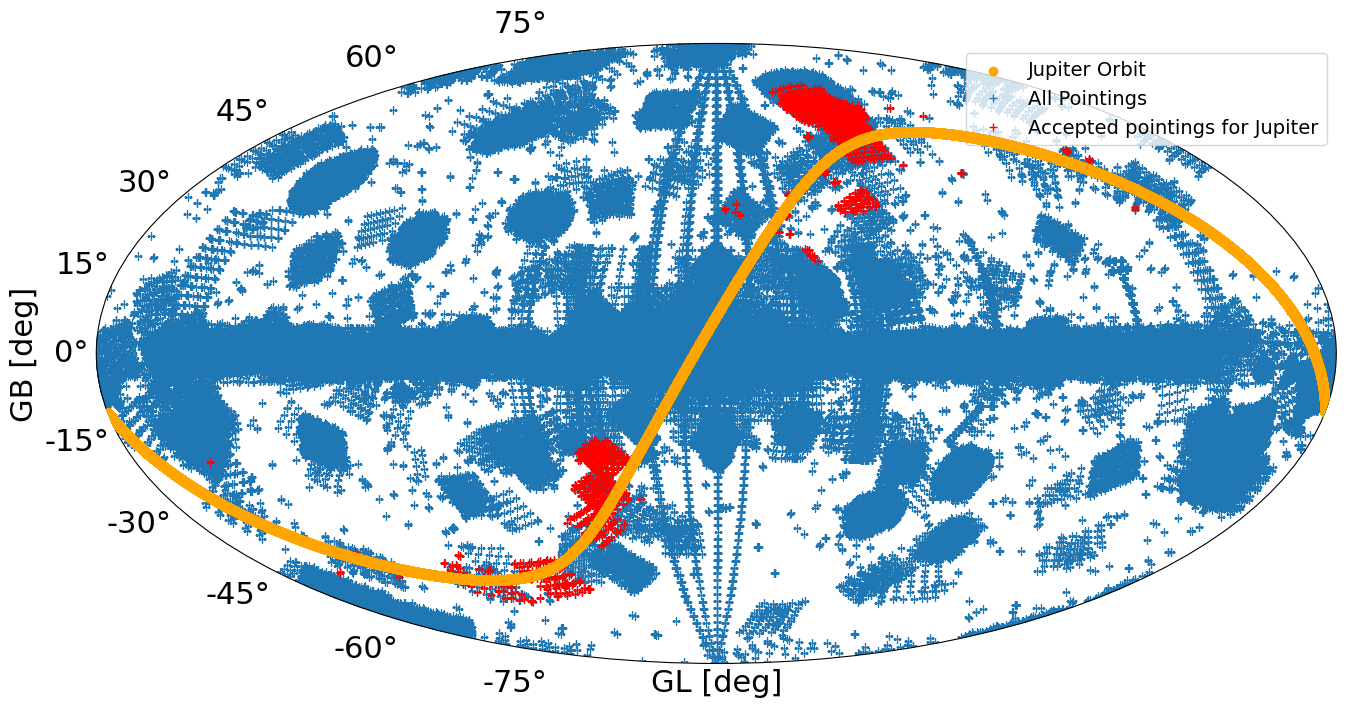}
    \caption{INTEGRAL/SPI observations within 22 years (blue crosses) and Jupiter orbit (orange). Chosen observations (after removing the  Center and Anti-center) which had Jupiter in the field of view are marked in red. The fully coded field of view of SPI is approximately 16$^\circ$ in diameter.}
    \label{fig:Pointings}
\end{figure}

INTEGRAL has been operating for 22 years, observing phenomena outside of the Solar System, while the Solar System itself has received little to no attention, except for the study of solar flares \citep{KienerFlares}. The INTEGRAL observation schedule was planned according to competing proposals, so that only serendipitous observations of Solar System objects had been performed, except for the targeted observation of Jupiter and Saturn in conjunction in the year. The total observation time of INTEGRAL during the 22 years is about 400 Ms in more than 200000 pointed observations ("pointings") of typically 2000 s each. For our analysis, we first select all pointings that coincide temporally and spatially with the location of Jupiter. Information regarding the orbit of Jupiter as seen from Earth was obtained using the Jet Propulsion Laboratory's (JPL) astroquery tool called Horizons \citep{JPLHorizons}. This results in 4625 pointings. We remove the Galactic Centre and anti-centre, because of strong and variable sources in these regions, which would lead to difficulties in disentangling the expectantly much weaker Jupiter emission. We removed all pointings that are located 40$^\circ$ around the centre and the anti-centre. This can be seen in Fig.\,\ref{fig:Pointings}. One Jupiter orbit takes 11.85\,yr to complete, thus Fig.\,\ref{fig:Pointings} corresponds to almost
two full orbits of Jupiter. Our final selection consists of 1527 pointings with a total observation time of 4.08\,Ms.

\subsection{Flux extraction}
The SPI data analysis consists of separating the source flux from the background and fitting  a spectral model to the result to obtain the optimized model parameters. The data themselves are structured as multi-dimensional histograms, including the dimensions of energy, pointing(=time), and detector. The appropriate statistics for counting photons in these intervals are the Poisson statistics. The general expression of a Poisson distribution can be expressed as: 
\begin{equation}\label{Poisson}
    P(d| m) = \frac{m^d e^{-m}}{d!},
\end{equation}
where $ P(d| m)$ is the probability of observing d events in a given interval and m is the expected number of events. As mentioned  above, we use Maximum Likelihood Estimation to obtain the optimal model parameters for our measurements. This is done by maximizing the likelihood function given by
\begin{equation}\label{Likelihood}
    L(\theta; d) = P(d | \theta) = \prod_{i=1}^{N} P(d_i | \theta),
\end{equation}
where $\theta$ are the model parameters, $P(X | \theta)$ is the probability function to get the observed data given the parameters $\theta$ of the specific model of interest, $d_i$ is the $i$-th data point and N is the total number of data points.

In the energy range SPI is operating, one of the major issues that appears is that the measured $\gamma$-ray spectra are dominated by instrumental background (BG). This BG can either be discrete (line) radiation, which originates from nuclear de-excitation reactions, or continuum processes, such as bremsstrahlung \citep{BGRSPDB}. Both of these occur due to the exposure of the instruments and satellite material to cosmic rays. This BG is altering the measured spectra and there is no independent BG model available, that can be used to handle SPI measurements in a straight-forward way, such as from simulations \citep{Weidenspointner} or on-off measurements \citep{Li}. In order to construct a self-consistent BG model for SPI data analysis, we utilized the method described in \citet{siegert2019background}. With that we can perform likelihood calculations based on the model parameters and our data. For illustrative purposes, we denote the data by $d_i$, where the index $i$ runs over the independent parameters required to describe each data point. We express the data as a linear combination of sky models $S_i$ and background models $b_i$, yielding:
\begin{equation}
    d_i \sim m_i = \alpha S_i + \beta b_i,
\end{equation}
where the parameters $\theta$ here are given by $\alpha$ and $\beta$, which are optimized in the fit and $m_i$ is the model used to describe the data. In the case of our Jupiter dataset, we found that for the single events (SE) energy range ($30-600$ keV) 410 BG parameters are required. SEs correspond to events in which energy is deposited in only one of SPI's germanium (Ge) detectors. For pulse-shape-discriminated (PSD) events, which are single-detector events additionally registered by the pulse-shape-discrimination electronics operating in parallel with the main electronics, 112 BG parameters are required in the ($600-2000$ keV) range. Finally, for events registered by the high-energy (HE) electronics chain in the ($2-8$ MeV) range, the optimal number of BG parameters is 154. Furthermore, we used 34 point sources, which coincide with the ecliptic and our chosen dataset. This is crucial in identifying areas where there are discrepancies between model and data, thus allowing us to gauge the areas where the model needs adjusting. Data analysis is performed within the Offline Scientific Analysis (OSA) software package, provided by the INTEGRAL Science Data Center (ISDC) \citep{OSA2003}. We used the \texttt{spimodfit} \citep{Strong2005,Halloin2009} to extract the fluxes of the five stationary sources (3C 273, NGC 4388, NGC 5252, NGC 4593, 510 IGR J15161-382), and the moving target of Jupiter, plus the background.

At this point the analysis would be enough for a stationary source, but Jupiter, as a planet, moves continuously along the ecliptic and changes its sky position throughout the mission lifetime, with an apparent motion on the sky, of order $0.1-0.3^\circ$ per day. The timescale typically separating successive pointings in the dataset, due to its serendipitous nature, is a few days to months. Given SPI's coarse angular resolution ($\approx 2.7^\circ$) and limited field of view, this accumulated displacement is not negligible. Because coded-mask imaging reconstructs a source's flux by deconvolving the shadow pattern for an assumed sky position, treating Jupiter as static causes signal from different pointings to be reconstructed, at slightly different assumed positions rather than Jupiter's true location. This would reduce the sensitivity and lead to an incorrect instrumental response, as it depends on the source position.
%

To account for this, we modified our analysis to treat Jupiter as a spatially-dependent point source. For each pointing, we determined the sky position of Jupiter as seen from Earth. Then, we generate an individual source image for every pointing, in which Jupiter occupies the corresponding instantaneous sky coordinates. Then we convolve these images with the instrument response function (IRF) representative of the specific pointing geometry and detector configuration for each measurement.

\begin{figure}[h!]
    \centering
    \includegraphics[width=\linewidth]{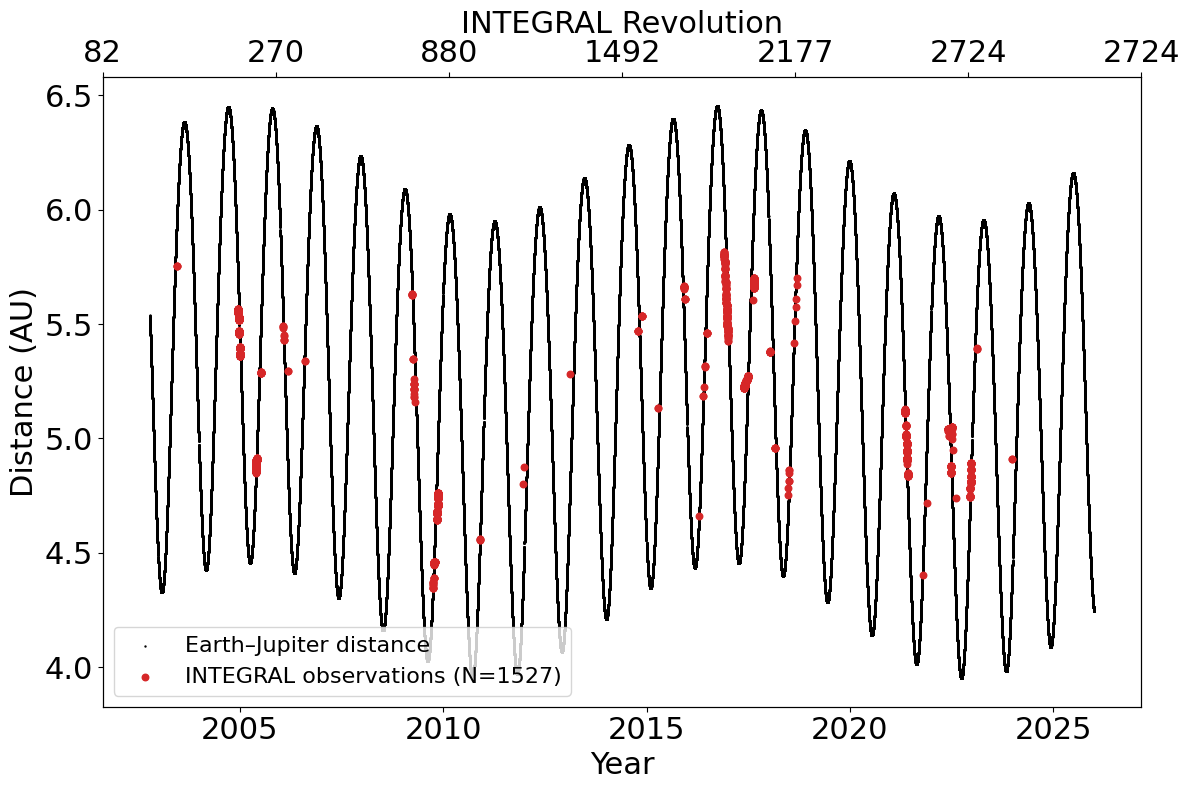}
    \caption{Distance between Earth and Jupiter in Astronomical Units from 2002 to 2025 shown in black. In red we provide the time where INTEGRAL/SPI observed Jupiter.}
    \label{fig:earth_jup}
\end{figure}

Because the two planets have different orbital periods, the separation between Jupiter and Earth varies. This leads to variations in  the expected counts, compared to the assumption that Jupiter is a fixed source, following the inverse square law. Earth orbits faster than Jupiter, so we have a minimum and maximum distance achieved periodically (typically between 4--6.5 \, AU), see Fig.\,\ref{fig:earth_jup}. To account for this, we need to adjust the expected fluxes by the proper distance between Earth and Jupiter, scaling them with the corresponding distance squared for each measurement, so that
\begin{equation} \label{earth-jup}
    C_i = \frac{\frac{1}{d{(i)}^2}}{\frac{1}{N}\sum_{j=1}^N\frac{1}{d(j)^2}}. 
\end{equation}
Here, $C_i$ is the correction factor applied to pointing $i$, $d_i$ is the counts in detector $i$, N is the total number of pointings. The denominator as a whole is the average inverse-square distance across the entire dataset. The variation in the separation between Earth and Jupiter is shown in Fig.\,\ref{fig:earth_jup}. In the same figure, we present the distance between the two planets at the times when INTEGRAL observed Jupiter.

\subsection{Spectral analysis method}
To constrain the physical properties of Jupiter's $\gamma$-ray emission, we performed spectral fitting using the Multi-Mission Maximum Likelihood framework \texttt{3ML} \citep {vianello20153ML}. This framework allows us to fit spectral models to the extracted flux measurements using both maximum likelihood estimation and Bayesian inference methods.

The differential photon flux incident on the Jovian atmosphere is not measured directly. What SPI records is a count spectrum that results from two successive transformations of the incident flux: the interaction
of the incident radiation with the atmosphere of Jupiter, and the response of the instrument. We adopt a forward-folding method, where a parametric source model is propagated through both transformations and compared to the data in count space. 

The observed emission is modelled as the sum of five physically distinct components, indexed by $c \in \mathcal{C} = \{\mathrm{CGB},\,p,\,e^{-},\,e^{+},\,\alpha\}$: the reflected CGB and four cosmic-ray-induced contributions from primary protons, electrons, positrons, and $\alpha$ particles. Each component carries its own incident spectrum $S^{(c)}(E)$ and its own atmospheric response, since the physics relating the incident particle to the emergent photon field differs from one component to the next.

Throughout this section we use two energy grids. The incident grid partitions the energy axis into $N_{\mathrm{in}}$ bins of edges $[E_i^{-}, E_i^{+}]$ and width $\Delta E_i = E_i^{+}-E_i^{-}$, sampling the spectrum arriving at the planet. The emergent grid, with bins $[E_j^{-}, E_j^{+}]$, samples the photon flux leaving the atmosphere, and the measured grid, with bins
$[E_k^{-}, E_k^{+}]$, samples the reconstructed photon energy at the detector.

For a single component $c$, the atmosphere relates the incident spectrum $S^{(c)}(E)$ to the photon flux emerging towards the observer through an atmospheric kernel $A^{(c)}(E' \mid E)$. The emergent photon flux is the convolution
\begin{equation}
  \label{eq:atm_cont}
  \Phi^{(c)}(E_j) =
  \int_{0}^{\infty} A^{(c)}(E_j \mid E_i)\, S^{(c)}_{i}\, \mathrm{d}E_i .
\end{equation}

For the CGB component, the incident and emerging particles are both photons and $A^{(\mathrm{CGB})}(E_j\mid E_i)$ is the energy redistribution probability density. For the four CR components the incident component is a particle, while the detected particle is a secondary photon, and $A^{(CR)}(E_j\mid E_i)$ is instead a production kernel, meaning the number of observable photons of energy $E_j$ generated per unit incident particle flux at energy $E_i$. The output of our Monte Carlo simulations (see Sect.\,\ref{sec:simulations}) for the yields of Jupiter's atmosphere is intrinsically discrete. It records, for initially injected particles and photons within a finite incident bin, the distribution of emergent photon energies in finite emergent bins.

To match this discretisation, we integrate Eq.~(\ref{eq:atm_cont}) over an emergent bin $j$. Assuming that the incident spectrum varies slowly across each incident bin, so that $S^{(c)}(E_i)\simeq S^{(c)}_i$ for $E\in[E_i^{-},E_i^{+}]$, the integral over the incident energy separates into a sum over incident bins,
\begin{equation}
\begin{aligned}
  \label{eq:atm_binned}
  \Phi^{(c)}_j
  &\equiv \int_{E_j^{-}}^{E_j^{+}}\!\!\Phi^{(c)}(E_j)\,\mathrm{d}E_j\\
  & = \sum_{i=1}^{N_{\mathrm{in}}}
    \left[ \int_{E_j^{-}}^{E_j^{+}}\!\!\!\int_{E_i^{-}}^{E_i^{+}}
    A^{(c)}(E_j\mid E_i)\,\mathrm{d}E_i\,\mathrm{d}E_j \right] S^{(c)}_i .
\end{aligned}
\end{equation}

The bracketed double integral is precisely the quantity obtained during the Monte Carlo simulation. Defining the photon yield matrix of component $c$ as
\begin{equation}
  \label{eq:atm_matrix_def}
  A^{(c)}_{ji} \equiv
  \int_{E_j^{-}}^{E_j^{+}}\!\!\!\int_{E_i^{-}}^{E_i^{+}}
  A^{(c)}(E_j\mid E_i)\,\mathrm{d}E_i\,\mathrm{d}E_j 
\end{equation}
 
Eq.~(\ref{eq:atm_binned}) reduces to the matrix-vector product
\begin{equation}
  \label{eq:atm}
  \Phi^{(c)}_j = \sum_{i=1}^{N_{\mathrm{in}}} A^{(c)}_{ji}\, S^{(c)}_i.
\end{equation}

For the CGB component $A^{(\mathrm{CGB})}_{ji}$ is the probability that an incident photon in bin $i$ is reflected in the emergent bin $j$. For a cosmic-ray component $A^{(c)}_{ji}$ is the mean number of observable
photons in bin $j$ produced per incident primary in bin $i$.

The emergent photon flux is folded through the SPI instrument response. Since no matter the input photon or particle the emergent one is always the same (photons at some energy), the IRF works the same for all cases and a single instrument kernel $R(E_k \mid E_j)$ applies. From this we get the expected count rate registered at measured energy $E_k$ per unit incident photon flux at energy $E_j$,
\begin{equation}
  \label{eq:irf_cont}
  C^{(c)}(E_k) =
  \int_{0}^{\infty} R(E_k \mid E_j)\, \Phi^{(c)}(E_j)\, \mathrm{d}E_j .
\end{equation}

Binning Eq.~(\ref{eq:irf_cont}) over the measured grid in
the same way as above yields
\begin{equation}
  \label{eq:irf}
  C^{(c)}_k = \sum_{j} R_{kj}\, \Phi^{(c)}_j ,
  \qquad
  R_{kj} \equiv \int_{E_k^{-}}^{E_k^{+}}\!\!\!\int_{E_j^{-}}^{E_j^{+}}
  R(E_k'\mid E_j)\,\mathrm{d}E_j\,\mathrm{d}E_k' .
\end{equation}

In practice $R_{kj}$ is the standard SPI redistribution matrix, provided in the OGIP response format and used without further processing. 

\subsection{Cosmic Background Radiation}
\subsubsection{CGB Photons}
 For the reflected CGB component, we adopt the smoothly broken power-law parametrization of \citet{Ajello_2008},

\begin{equation}
   \frac{dN}{dE} = \frac{C}{(E/E_B)^{\Gamma_1} + (E/E_B)^{\Gamma_2}},
\end{equation}

where $\Gamma_1$ is the powerlaw  index for low energies and $\Gamma_2$ the index for high energy. For these we use the value from \citet{Ajello_2008}, namely $\Gamma_1 = 1.3$ and $\Gamma_2 = 2.8$. We also take the value for $E_{B} = 30.0$ keV, which is the break energy. We leave the normalization $C$ as a free parameter, which we fit to the SPI data.
\subsubsection{Cosmic Rays}
Regarding the CR spectra we use the models and values described in \citet{Boschini_2018} for electrons and positrons, in particular,
\begin{equation}
F(T) =
\begin{cases}
\dfrac{1.181 \times 10^{11} \, T^{-12.061}}
      {1 + 4.307 \times 10^{8} \, T^{-9.269} + 3.125 \times 10^{-8} \, T^{-10.697}}, \\
\hfill T < 6.88~\text{GeV}, \\[1.5ex]
995.598 \, T^{-3.505} + 4.423 \, T^{-2.620}, \\
\hfill T \geq 6.88~\text{GeV},
\end{cases}
\label{eq:boschini_electron}
\end{equation}

while, in the case of protons and $\alpha$ particles we use the equation found in \citet{Boschini_2020},
\begin{equation}
F_{\rm H}(R) =
\begin{cases}
aR^2 + bL(R)^2 + (c + d\tilde{R})\, G\!\left(\dfrac{R}{L(R)}\right) - f - g\tilde{R}  \\
 \quad - hL(R)\, i^{\, G\left[\frac{R}{L(R)}\right]} , \\
\hfill R \leq 2.5~\text{GV}, \\[1.5ex]
R^{-2.7}\!\left[ -l - mR + nL(R) + oG(pR) \right. \\
\left. \quad + (qR - s^{L(R)} r) \cos(t + u\tilde{R}) \right], \\
\hfill R > 2.5~\text{GV}.
\end{cases}
\label{eq:boschini_proton}
\end{equation}

The parameter values can be found in the same work. We combine the different CR particle models into a single template and introduce a multiplicative overall scaling factor $M$, which serves as the amplitude and is fitted to the data. Deviations of $M$ from unity quantify the strength of a potential signal, with larger values indicating that a stronger signal would have been required for detection. 

\section{Intrinsic Jupiter $\gamma$-ray Emission }\label{DM}
\subsection{Hard x-ray estimates}
Jupiter's intrinsic hard X-ray emission is a result of bremsstrahlung emission, arising from the precipitation of energetic electrons into the upper atmosphere, where they decelerate and emit a power-law continuum. This was first detected with XMM-Newton in the 2–7 keV band, and  it accounts for essentially all auroral flux above 2 keV with a hard spectral index \citep
{XMMJupiter}. The hard component is well described by a power-law with index $\Gamma \sim 0.2$, while the soft continuum below 2 keV is fitted by a thermal bremsstrahlung model with $kT \sim \mathrm{0.1-0.3\,keV}$. A significant part of that originates from Io's plasma torus, where the  X-ray emission from this electron population was identified by \citet{Elsner_2002}.
NuSTAR observations subsequently confirmed and extended this picture, detecting hard X-ray emission from both auroral regions in the $\mathrm{8 - 20 \,keV}$ band with a flat non-thermal power-law slope of $\Gamma = 0.60 \pm 0.22$ \citep{Mori2022}. The broad-band 3–20 keV spectrum from NuSTAR is characterized by a spectral cutoff above $\sim20\mathrm{\,keV}$. Over the $3$--$20$~keV range, \citep{Mori2022} measured a flux at Earth of $3.3^{+0.2}_{-0.8} \times 10^{-6}\,\mathrm{ph\,cm^{-2}\,s^{-1}}$, while their bremsstrahlung simulations based on the precipitating electron population predicted a flux at Earth of $9.1 \times 10^{-7}\,\mathrm{ph\,cm^{-2}\,s^{-1}}$. Moreover, the Juno/JEDI observations directly characterize the precipitating electron population responsible for this emission, exhibiting power-law-like energy spectra extending well beyond 800 keV, with downward energy fluxes reaching up to $\mathrm{280\,mW\,m^{-2}}$ \citep{Mauk2017}. The fraction of precipitated electron energy converted into bremsstrahlung photons scales as
\begin{equation}
\frac{F_{\gamma}}{F_e} \sim Z\,2\, \alpha \frac{E_\gamma}{E_e},    
\end{equation}
 from \citet{Longair}, where $\alpha$=1/137 is the fine structure constant and Z=1 for the hydrogen-dominated Jovian atmosphere. Applying this to the results of \citet{Mauk2017}, the bremsstrahlung surface brightness above 20 keV is of order $\mathrm{10^{-5} - 10^{-6}\,erg\,cm^{-2}\,s^{-1}\,sr^{-1}}$. Since bremsstrahlung emission is produced by precipitating electrons impacting the atmosphere, it is expected to be concentrated in the auroral regions. The precipitating electrons follow Jupiter's magnetic field lines and therefore enter the atmosphere preferentially near the magnetic poles, where the auroral ovals are located. We thus assume that the bremsstrahlung emission is confined to the auroral oval rather than distributed over the full Jovian disk, and we scale the local emission rate by the oval area of $ \mathrm{10^{17}\, cm^2}$ \citep{ovalArea}. For the average distance between Earth-Jupiter of 5.2 AU, we find that the expected bremsstrahlung flux at Earth would be of the order $\mathrm{10^{-10}\,ph\,cm^{-2}\,s^{-1}\,keV^{-1}}$ at 20 keV.

\subsection{Dark Matter Model}
The model we consider here produces $\gamma$-rays through the annihilation of dark matter and subsequent decay of the mediator \citep{JupiterDM}. In this scenario, weakly interacting particles may scatter with atmospheric or interior material, lose kinetic energy, and become gravitationally bound, leading to an enhanced local dark matter density inside Jupiter. The resulting $\gamma$-ray spectrum depends on the dark matter particle mass and annihilation cross-section. The emission observed from Jupiter can constrain these parameters in the SPI MeV region.
The differential $\gamma$-ray flux predicted by the model is given by

\begin{equation}
E^{2}\frac{d\Phi}{dE}
=
\frac{\Gamma_{\rm ann}}{4\pi D_{\oplus}^{2}}
\times
E_{\gamma}^{2}\frac{dN_{\gamma}}{dE_{\gamma}}
\times
\mathrm{BR}(X \rightarrow \mathrm{SM})
\times
P_{\rm surv},
\end{equation}

following \citet{JupiterDM}. Here, $D_{\oplus}$ is the average
distance between Earth and Jupiter, namely 5.2 AU,
$\mathrm{BR}(X \rightarrow \mathrm{SM})$ is the branching ratio
of the mediator into a given Standard Model final state,
and $P_{\rm surv}$ is the probability that the mediator escapes
Jupiter and decays before reaching Earth.
The relevant annihilation channel is
\begin{equation}
\chi\bar{\chi} \rightarrow \phi\phi \rightarrow 4\,\mathrm{SM},
\end{equation}
which produces a box-shaped $\gamma$-ray spectrum. Following
\citet{SolarDM}, the spectrum is parameterized as
\begin{equation}
\frac{dN_\gamma}{dE_\gamma}
=
\frac{N_\gamma}{\Delta E}
\Theta(E_\gamma - E_-)
\Theta(E_+ - E_\gamma),
\end{equation}
where $N_\gamma$ is the number of photons produced per annihilation,
$\Theta$ is the Heaviside step function, and the spectral width, with dark matter particle mass being $m_x$ and mediator mass $m_y$, is
\begin{equation}
\Delta E
=
m_X
\sqrt{1-\frac{m_Y^2}{m_X^2}}.
\end{equation}

The endpoints of the box spectrum are
\begin{equation}
E_{\pm}
=
\frac{m_X}{2}
\left(
1 \pm
\sqrt{1-\frac{m_Y^2}{m_X^2}}
\right).
\end{equation}

The mediator survival probability is given by
\begin{equation}
P_{\rm surv}
=
e^{-R_J/(\gamma c\tau)}
-
e^{-D_\oplus/(\gamma c\tau)},
\end{equation}
where $R_J$ is the radius of Jupiter,
$\tau$ is the mediator lifetime,
and $\gamma$ is the Lorentz boost factor.

The normalization of the predicted flux is determined by the
dark matter annihilation rate
\begin{equation}\label{anihh}
\Gamma_{\rm ann} = \frac{C_{\rm tot}}{2},
\end{equation}
assuming equilibrium between dark matter capture and annihilation
inside Jupiter. At equilibrium, the number of dark matter particles captured per unit time equals the number removed through annihilation. Since each annihilation involves two dark matter particles, the annihilation rate is half the total capture rate, giving the factor of $1/2$ seen in Eq.\,\ref{anihh}. The total capture rate $C_{\rm tot}$ depends on
the optical depth,
\begin{equation}
\tau = \frac{3}{2}\frac{\sigma}{\sigma_{\rm sat}},
\end{equation}
where $\sigma$ is the dark matter--nucleon scattering cross section
and $\sigma_{\rm sat}$ is the saturation cross section for capture. Here $\sigma$ is treated as the free parameter and is constrained through a fit to the SPI data. The full expression for $C_{\rm tot}$ follows the multiscatter
capture formalism described in \citet{JupiterDM}, along with the values for the parameters used.

\section{Atmospheric Simulations}\label{sec:simulations}
\subsection{Atmospheric Model}
Alongside the process described above, we simulated the atmospheric yield of Jupiter, when bombarded with high energy photons and cosmic rays. We used $\texttt{MEGAlib}$ \citep{Megalib}, which contains $\texttt{Geomega}$ and $\texttt{Cosima}$, which are used to create 3D geometry and material models of satellites and instruments. Here, we use the $\texttt{cosi-atmosphere}$ plug-in \citep{cosi} to model Jupiter's atmosphere. The pipeline was initially constructed to simulate the scattering of photons within the atmosphere of Earth, since when $\gamma$-rays travel through the atmosphere they may undergo Compton scattering, causing a distortion of the signal. This is important for high-altitude balloon observations. We adapted the \texttt{cosi-atmosphere} code for Jupiter by replacing the atmospheric composition of Earth with that measured by the Galileo probe \citep{Atmo} and adjusting the relevant simulation parameters, such as the radius of the planet, atmospheric intervals in which the composition is defined, and the surrounding sphere radius from which the simulated particles originate. We simulated the atmospheric yield of Jupiter to the incoming high energy photons and particles. 

\begin{figure}[h!]
    \centering
    \includegraphics[width=\linewidth]{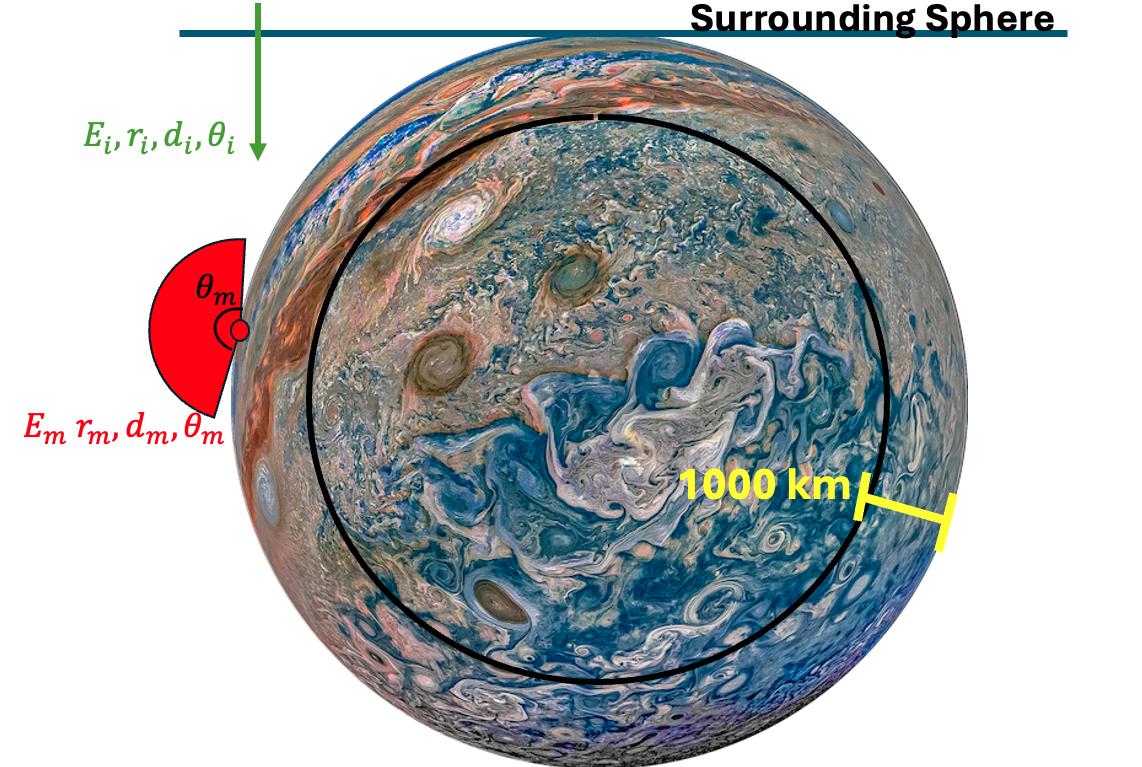} 
    \caption{The diagram illustrates the simulation setup. In red is shown the measured energy, position, direction, and angle of outgoing photons. In green, the initial properties of the simulated photons are displayed, including their energy, position, direction, and angle. The total atmospheric depth considered is displayed in yellow. Concentric rings represent the maximum atmospheric depth (inner ring) and some location in the atmosphere within our study range. We note that the scale is illustrative only. The surrounding sphere denotes the region from which $10^6$ photons are emitted isotropically around Jupiter’s atmosphere. The scheme is on top of an image of Jupiter \citep{Jupiter}.}
    \label{fig:sim_scheme}
\end{figure}

\begin{figure}[h!]
    \centering
    \includegraphics[width=\linewidth]{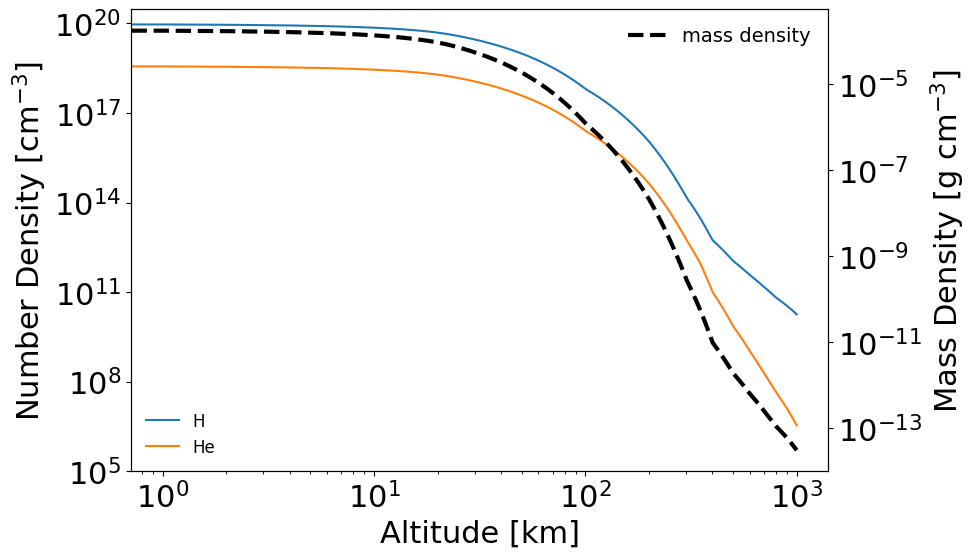}
    \caption{Jupiter's atmospheric model, as measured from the Galileo probe \citep{Atmo}. Shown are only the relevant components for number (left) and mass (right) density. Molecules, such as $\mathrm{CH_4}$, do not significantly contribute.}
    \label{fig:model}
\end{figure}

In our case, we were interested in the photons that escape the atmosphere after scattering, as these will be the ones potentially measurable by MeV telescopes. That means we created a
spherical model of Jupiter’s atmosphere, composed of spherical shells of thickness of 1 km each, reaching into Jupiter down to 1000 km. Using data from \citet{Atmo} the calculated composition model gives us the pressure ($p$), the temperature ($T$), the mean molecular weight ($\mu$) and the mole fraction of each component for intervals of 100 km. Using that, we calculated the the number density of the atmospheric components and the total mass density per altitude. We first calculate the specific gas constant $R_{S}$ for the gas mixture, using the universal gas constant ($R$) and the mean molecular weight ($\mu$) using

\begin{equation} \label{Rspec}
    R_\text{specific} = \frac{R}{\mu}.
\end{equation}
With that we calculated the mass density $\rho$ using the ideal gas law rearranged to solve for the density

\begin{equation} \label{density}
    \rho= \frac{p}{R_\text{specific}T}.
\end{equation}
Finally, the number densities for atomic H and He were calculated using the molar fractions

\begin{equation} \label{density}
    n_i= \frac{pf_i}{M_i},
\end{equation}
where $\rho$ is the mass density, $f_i$ is the molar fraction of element $i$ and $M_i$ is the molar mass of $i$. The values were then logarithmically interpolated, to obtain the mass density and the number density for
intervals of 1 km. We show a scheme of the simulated geometry in Fig.\,\ref{fig:sim_scheme}. The corresponding atmospheric model is shown in Fig.\,\ref{fig:model}.

\subsection{COSI-Atmopshere Setup}
We simulate an isotropic source with a powerlaw photon spectrum of the form $dN/dE \approx E^{-1}$, yielding a constant number of simulated photons per logarithmic energy interval. The range we utilise is from 10 keV to 100 MeV using $10^6$ initial photons. We also simulate the interactions of cosmic rays with the atmosphere. These are protons, electrons, and alpha particles, in the range of 1 MeV to 10 GeV, as well as, positrons in the range of 100 keV to 10 GeV. The source is simulated using a surrounding sphere with a radius of Jupiter plus 1000 km. To replicate plane waves hitting the
atmosphere, photons are emitted perpendicularly from a disk located on the sphere’s surface,
sharing the same radius. In the case of an isotropic source, photons are uniformly directed
towards Jupiter’s surface from all incident angles across the sky. The simulations capture both
the initial ($i$) and measured ($m$) values of the photon’s energy ($E$), position ($r$), and direction
($d$), as shown in Fig.\,\ref{fig:sim_scheme}. The measured values are obtained by tracking the photons (reflected and induced) whenever they escape the world volume, meaning they escape Jupiter's atmosphere. The flux measured from an observer at Earth is then scaled by the inverse distance squared, to obtain the observed flux.

\section{Results}\label{sec:results}
\subsection{Analysis results}
\begin{figure}[h!]
      \centering
      \includegraphics[width=\linewidth]{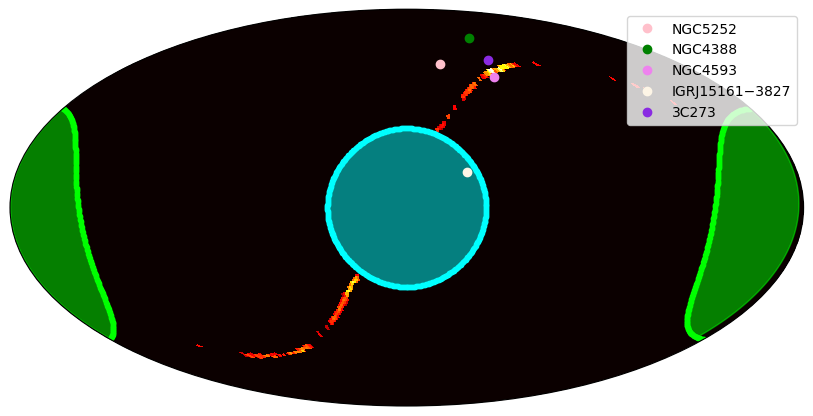}
      \caption{Positions where SPI observed Jupiter most often, after removing the Galactic center
(blue circle) and the Galactic anti-center (green area). The intensity goes from black (no observation) to white (maximum observation time). The colored dots represent the sources also in the field of view of SPI when observing Jupiter.}
      \label{fig:Jup_obs}
  \end{figure}

A heatmap of Jupiter's positions as observed with SPI is shown in Fig.\,\ref{fig:Jup_obs}. Each point marks Jupiter's position in the sky at a given observation time, and combining these positions across the dataset produces a heatmap of its most frequent locations. The map also shows the excluded regions in cyan (Galactic Centre) and green (Anticentre). The displayed sources indicate other sources that may lie within the SPI field of view during observations of Jupiter and therefore need to be taken into account in the analysis. Consequently, sources located within an excluded region, such as the source near the Galactic Centre, may still contribute to observations for which Jupiter lies close to the boundary of that region.

\begin{figure}[h!]
    \centering
    \includegraphics[width=\linewidth]{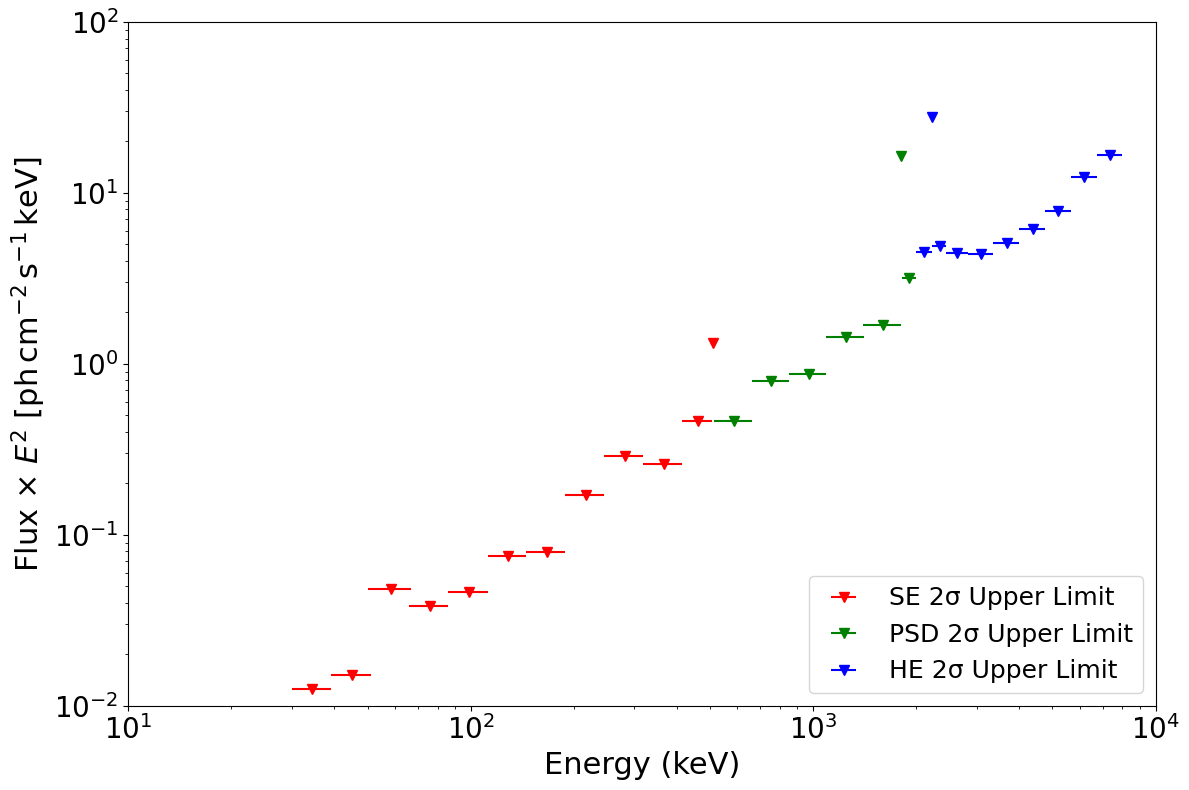}
    \caption{Extracted data points from SPI (2$\sigma$ upper limits) measurements.}
    \label{fig:results_analysis}
\end{figure}

Having the position of Jupiter for each SPI pointing is essential not only to identify the observations in which Jupiter was present in the field of view (fov), but also to generate the per-pointing sky images described in Sect.\,\ref{sec:data_analysis}, in which Jupiter is represented as a single pixel. These images were convolved with the IRF and fitted as a combined sky model, from which the flux of Jupiter was extracted with the background removed.
The flux extracted, by fitting the final distance-corrected expectations of the moving Jupiter source to the dataset with \texttt{spimodfit} \citep{OSA2003, Strong}, across the SPI energy range, is shown in Fig.\,\ref{fig:results_analysis}. The measurement is split across three detector event types. SE events, (30–600 keV), PSD events (600–2000 keV), and HE events (2–8 MeV), each requiring a separate background model due to the distinct detector response and background composition for these energy regimes. All measured values are consistent with zero within uncertainties, and therefore we report $2\sigma$ upper limits on the flux in each energy bin from 30\,keV to 8\,MeV. 

These upper limits represent, to our knowledge, the first spectral measurement of a planetary atmosphere in the MeV band with a $\gamma$-ray telescope other than Earth, and they establish the baseline sensitivity against which the physical models are compared. The limits span roughly three orders of magnitude in flux across the band, from $\approx 10^{-2}$ to $\approx 10^{1}$ $\mathrm{ph\,keV\,cm^{-2}\,s^{-1}}$ in a $\mathrm{Flux}\times E^2$ representation. The three higher upper limits correspond to narrow energy bins centered on prominent $\gamma$-ray lines: the 511 keV electron–positron annihilation line, the 1.8 MeV line from radioactive decay of $^{26}$Al, and the 2.2 MeV neutron-capture line. The narrower bin widths result in larger statistical uncertainties compared with the broader neighboring bins, causing the corresponding $2\sigma$ upper limits to appear higher.

\begin{table*}[t]
\caption{Spectral index and integrated flux of the five sources analysed.}
\label{tab:results}
\centering
\small
\begin{tabular}{lcccccc}
\toprule
\multirow{2}{*}{Source}
& \multicolumn{2}{c}{Maximum likelihood}
& \multicolumn{2}{c}{Bayesian inference}
& \multicolumn{2}{c}{Literature} \\
\cmidrule(lr){2-3}
\cmidrule(lr){4-5}
\cmidrule(lr){6-7}
& $\alpha$ & $F\,(10^{-6})$
& $\alpha$ & $F\,(10^{-6})$
& $\alpha$ & $F\,(10^{-6})$ \\
\midrule
3C\,273
& $-1.6 \pm 0.1$ & $1.2^{+0.9}_{-0.5}$
& $-1.6 \pm 0.1$ & $1.0^{+0.4}_{-0.3}$
& $-1.5$ & $1.1$ \\
NGC\,4388
& $-1.6 \pm 0.1$ & $7.3^{+5.2}_{-3.1}$
& $-1.5 \pm 0.1$ & $7.5^{+2.5}_{-4.4}$
& $-1.5$ & $0.9$ \\
NGC\,5252
& $-2.5 \pm 3.4$ & $2.4^{+0.7}_{-1.0}$
& $-5.9 \pm 2.5$ & $25.2^{+60}_{-60}$
& $-2.3$ & $3.9$ \\
NGC\,4593
& $-1.9 \pm 0.6$ & $1.1^{+9.0}_{-1.0}$
& $-1.7 \pm 0.2$ & $0.5^{+0.5}_{-0.3}$
& $-1.8$ & $1.4$ \\
IGR\,J15161$-$3827
& - & -
& - & -
& $-2.65$ & $9.44$ \\
\bottomrule
\end{tabular}
 \tablefoot{Spectral index and integrated flux (30--1000\,keV) from maximum likelihood, Bayesian inference, and literature values \citep{bouchet2008integral}. Uncertainties propagated via linear (first-order) error propagation.}
\end{table*}

\begin{figure}[h!]
    \centering
    \includegraphics[width=\linewidth]{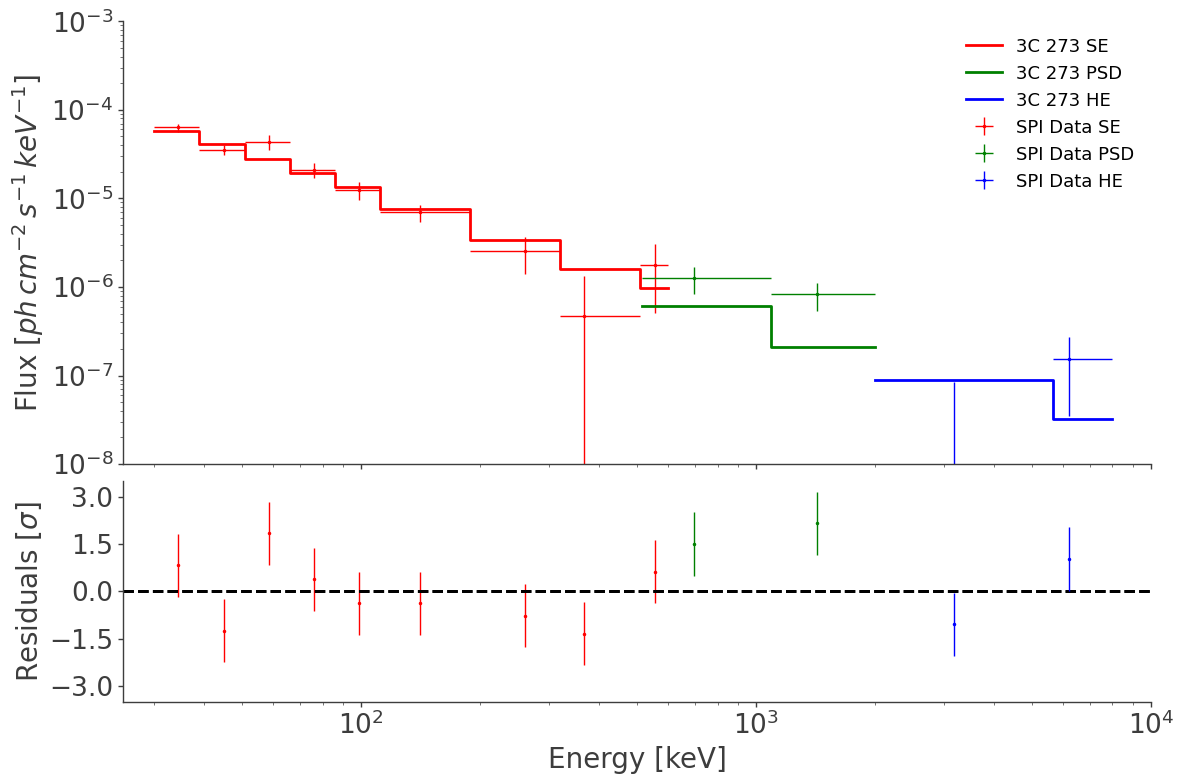}
    \caption{Top: Measured spectrum of 3C\,273 in our Jupiter-focussed dataset (crosses), together with a power-law fit (histogram), for the energy range of SPI. Single Events (SE) are shown in red crosses, Pulse Shape Discrimination (PSD) in blue crosses and High Energy events (HE) with green crosses. Bottom: Residuals of the fit.}
    \label{fig:3c273}
\end{figure}

 In total, we found five relevant sources, 3C\,273, NGC\,4388, NGC\,5252, NGC\,4593, IGR\,J15161-3827. These are sufficiently significant to be included in the analysis, as their emission may interfere with the very low count rates expected from Jupiter. Here, we show the resulting spectrum of 3C\,273 in Fig.\,\ref{fig:3c273}. We fit the spectrum with a power-law model. From that we calculated the integrated flux between $30-1000$ keV and compared it to values in the literature. To fit our data, we use both maximum
likelihood fits and Bayesian inference. Regarding Bayesian inference, we use a \texttt{uniform} prior for the index and \texttt{loguniform} prior for the amplitude. In Fig.\,\ref{fig:3c273} we present only the maximum likelihood result, but in Tab.\,\ref{tab:results} we present the values for all five sources. The power-law index ($\alpha$) and the amplitude ($K$) align with those reported in the existing literature \citep{Bouchet}. We note that NGC 5252 is a relatively weak source in the SPI data, resulting in poor constraints on its spectral shape and, consequently, large uncertainties in its spectral index and integrated flux. Overall agreement with previous measurements, suggests the validity of our methodology and supports the accuracy of the results obtained for Jupiter.

\subsection{Simulation results}
\begin{figure*}[!t]
    \centering
    \includegraphics[width=0.8\textwidth]{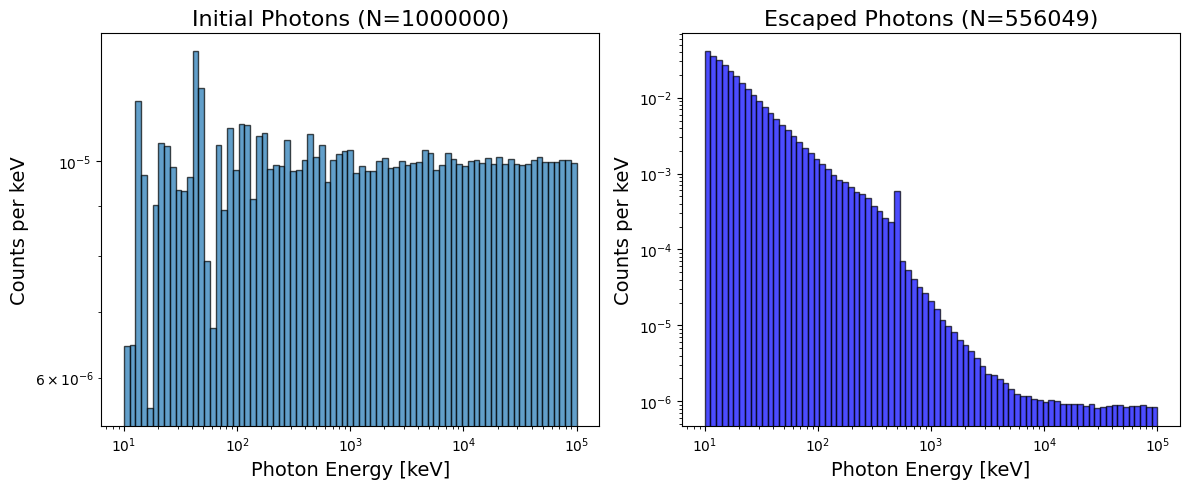}
    \caption{Photon distribution for initial (left) and backscattered photons (right) from 10\,keV to 10\,MeV, binned into 100 logarithmically spaced bins. The interaction of photons with Jupiter's atmosphere redistributes photons toward lower energies through down-scattering processes.}
    \label{fig:phtns_init_scatrd}
\end{figure*}
From the simulations of Jupiter's atmospheric yield, we can see in Fig.\,\ref{fig:phtns_init_scatrd}, the distribution of energies for the initial and backscattered photons for Jupiter, between 10 keV and 100 MeV. Here we use 100 logarithmically spaced energy bins. We can clearly see the expected higher flux of lower energy photons, caused by the scattering processes in the atmosphere such as Compton scattering, Rayleigh scattering, pair creation, and subsequent positron annihilation. We can distinctly see a line in the continuum, which can be attributed to the expected 511 keV line. This is expected since pair production has a threshold of 1022\,keV, well within the energy range studied here. This then, results in electron-positron annihilation, leading to the creation of 511 keV photons, and in the formation of positronium which may either decay into two photons of 511 keV each, or a rising continuum of ortho-positronium with three photons \citep{ortho}. Although the production of 511 keV photons is therefore expected, the simulation highlights that, potentially, if the 511 keV emission from Jupiter, other planets, or asteroids reaches levels comparable to those of known 511 keV sources, it could contribute to or bias Galactic and high-latitude observations.

\begin{figure}[h!]
    \centering
    \includegraphics[width=\linewidth]{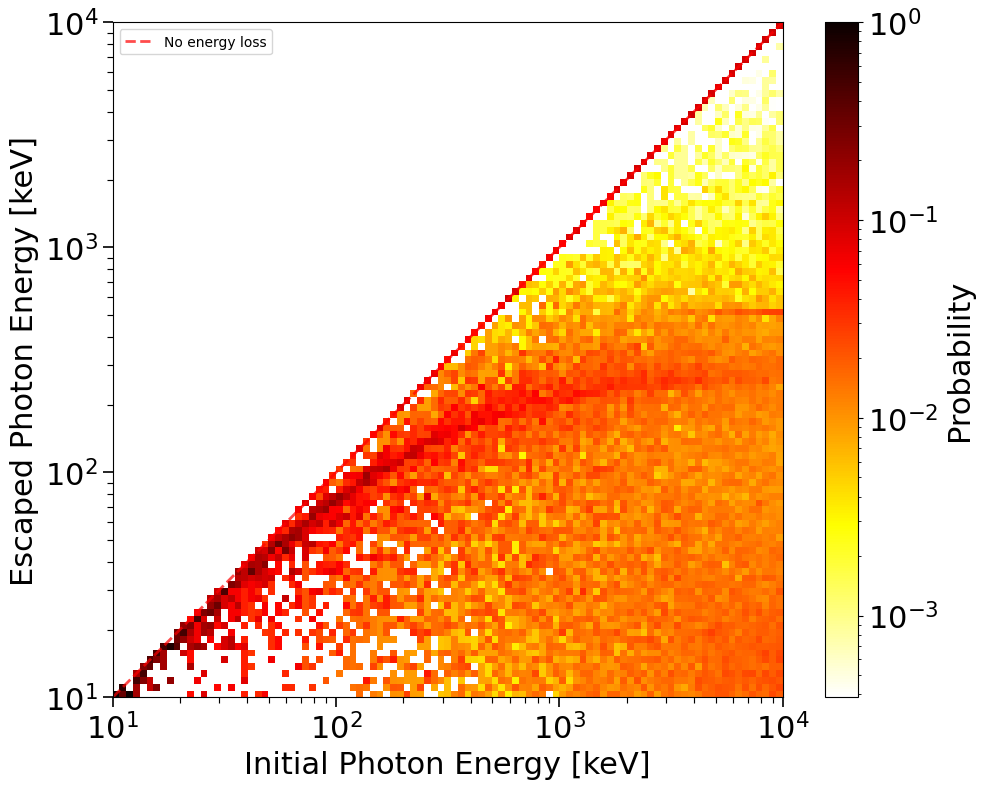}
    \caption{Response matrix of the interaction of incoming photons with energy $E_i$, with Jupiter's atmosphere. Here we see how the atmospheric interaction, the scattering of the incoming photons, changes the escaped photon. $E_i$ is the initial photon energy when entering the atmosphere and $E_m$ is the observed energy after it has been scattered. The values in the diagonal correspond to measured photons that have the same energy as the initial and thus penetrate the atmosphere without scattering, while off-diagonal values correspond to lower energy measured photons, that have been scattered. }
    \label{fig:matrix}
\end{figure}

We created yield matrices, which are projections of the initial and measured photon energies onto a graph with $E_i$ and $E_m$ energies. We bin those matrices using 100 logarithmically spaced bins. In Fig.\, \ref{fig:matrix} we show the response matrix for the simulation of incoming photons. The color scale gives the ratio of photons that are detected with a measured energy, $E_m$, given an initial energy, $E_i$. The diagonal values correspond
to measured photons that have the same energy as the initial
and thus penetrate the atmosphere without scattering. Off-diagonal values correspond to scattered photons. Here, the measured photons have an energy lower than the initial ones, due to their interaction with the atmosphere. Atmospheric interactions can only cause energy losses to the incoming photons, thus we do not measure any photons with energy higher than the initial one. This can then be applied to any input spectrum we want to study, and obtain the spectrum that would be measured by our instrument. The same kind of matrix can be generated for the interaction between CR and atmosphere, and in Appendix \ref{AppendixCR} we showcase the ones for protons, electrons, alpha particles and positrons.

\begin{figure}[h!]
        \centering
        \includegraphics[width=\linewidth]{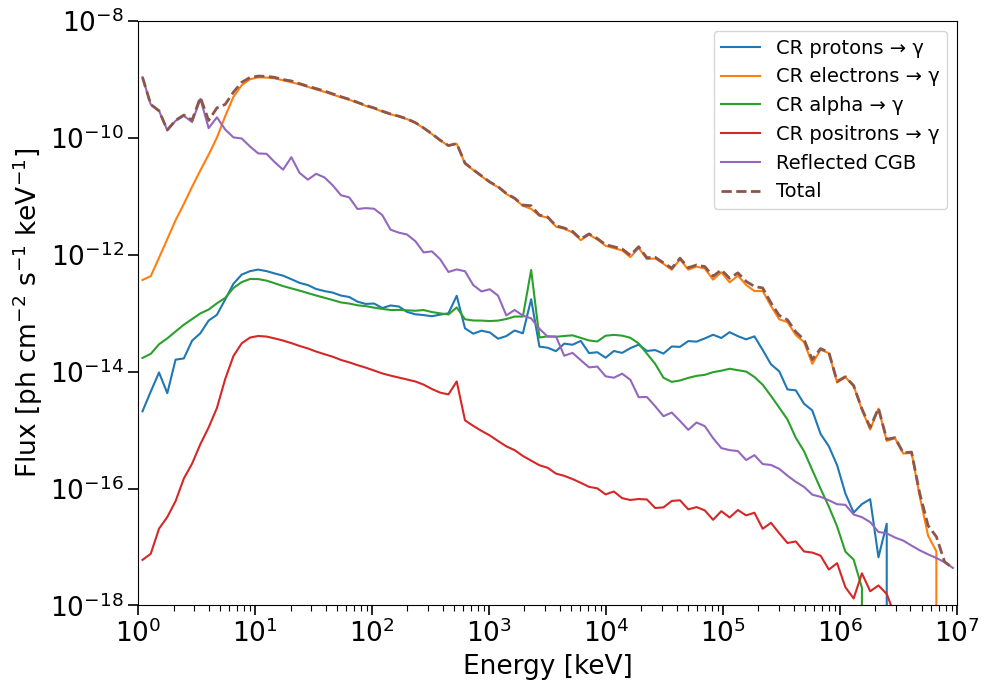}
        \caption{Photons measured at earth from the reflected CGB, as well as, the ones generated from CR-atmosphere interaction.}
        \label{fig:total_resp}
    \end{figure}

In Fig.\,\ref{fig:total_resp}, we show how the matrices applied to general CGB and CR spectra behave. Here, with the dashed line, we can see the total spectrum emitted from Jupiter as seen from a distance of $5.2\,\mathrm{AU}$ from the planet. The CR-induced proton and alpha components exhibit a feature around 2.2 MeV, corresponding to the 2.223 MeV neutron-capture line produced by secondary neutrons in the atmosphere. In contrast, the 511 keV annihilation feature is not prominent in the reflected CGB component, which is dominated by Compton-scattered incident CGB photons rather than secondary emission. The reflected CGB also dominates the predicted flux below \(\sim1\) MeV, due to the relatively high incident CGB flux and efficient Compton backscattering in this energy range. This can then be combined with our 3ML fitting to get constraints on the flux coming from Jupiter, as is described in section \ref{sec:data_analysis}. 

\section{Discussion} \label{Discussion}
\subsection{Comparison with the literature.}
To interpret the flux measured from Jupiter and offer theoretical constraints, we considered several models, each representing a distinct physical emission mechanism. These were: (i) a cutoff power law, representing generic non-thermal emission; (ii) a dark matter annihilation model; (iii) a reflected CGB spectrum, in which the incident isotropic background is scattered by Jupiter's atmosphere; and (iv) a particle-induced emission model, in which cosmic-rays impacting Jupiter's atmosphere produce $\gamma$-rays through bremsstrahlung and hadronic interactions, respectively \citep{Longair}. This enables a direct comparison of the derived upper limits with values reported in the literature, providing a measure of the consistency between our results and existing measurements. We compare our constraint to the value reported in \citet{Ajello_2008}.

\begin{figure}[h!]
    \centering
    \includegraphics[width=\linewidth]{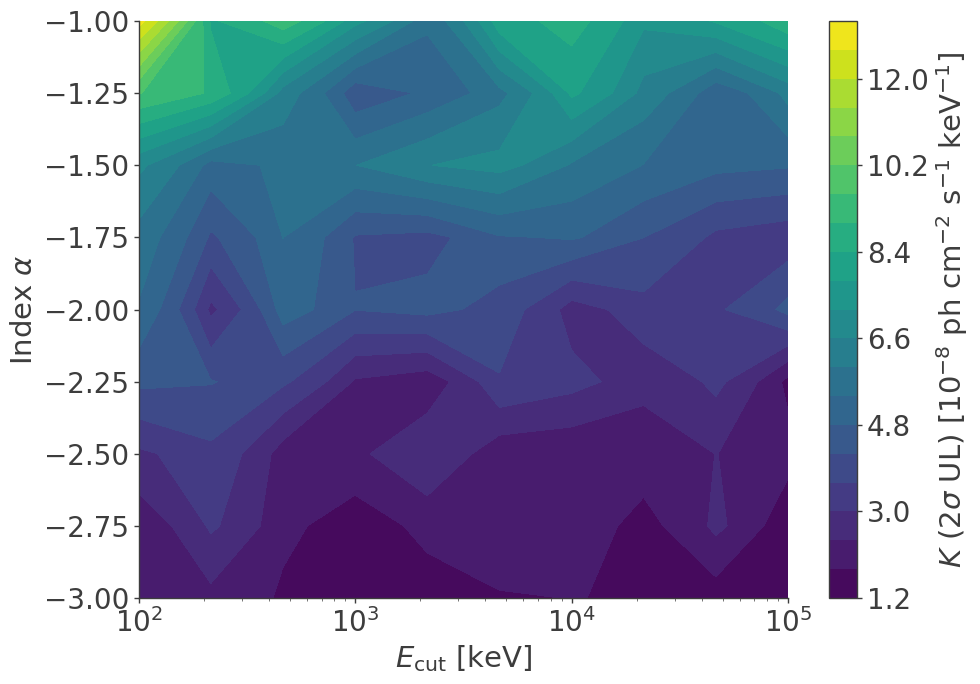}
    \caption{Upper limits on the cutoff power law amplitude K (Eq.\,\ref{cutoffpowerlaw}).}
    \label{fig:3C273}
\end{figure}

The cutoff power-law provides a flexible empirical description of the observed spectrum without assuming a specific emission mechanism, parametrized by a photon index, normalization, and high-energy cutoff. This model serves as a baseline for characterizing the spectral shape and deriving the total $\gamma$-ray luminosity, in the band between 30 and 8000 keV, from the data. The CGB illuminates Jupiter isotropically. A fraction of this emission is reflected, through various processes, back toward the observer, with the reflected spectrum modified by the energy-dependent scattering cross-section and atmospheric opacity. Unlike the CR induced component, the CGB reflection amplitude depends only on the geometric solid angle subtended by Jupiter and the incident CGB intensity, varying solely with distance to the Earth, rather than solar activity. This distance also plays a role in the CR induced flux, but is subdominant. Additionally, the dark matter model we consider here produces $\gamma$-rays through the annihilation of dark matter and subsequent decay of the mediator \citep{JupiterDM}. Incident spectra, based on models that we described in Sect.\,\ref{sec:simulations}, are forward-folded through the energy-dependent atmospheric yield matrix to predict the reflected spectrum observed at Earth. In the case of the cutoff powerlaw, the free parameters in this model are an overall amplitude scaling and the spectral index. The cutoff powerlaw used is described by
\begin{equation}\label{cutoffpowerlaw}
\frac{dN}{dE} = K \left(\frac{E}{E_0}\right)^{\alpha} \exp\left(-\frac{E}{E_{\rm cut}}\right),
\end{equation}
where $K$ is the amplitude, $E_{cut}$ is the cutoff energy, $E_0$ is the pivot energy, and $\alpha$ is the spectral index. We perform the spectral fitting using a Bayesian framework with the 3ML software package, employing MultiNest \citep{Multinest} for parameter estimation and evidence calculation. In Fig.\,\ref{fig:3C273} we show the log-likelihood plot for the powerlaw index versus the cutoff energy, as well as the same plot for the upper limit of 2$\sigma$, for the amplitude K, for a powerlaw model, as a function of cutoff energy and index. To produce these plots, we scan the powerlaw index $\alpha$ from -3.0 to -1.0 in steps of 0.25 and the cutoff energy $E_{\rm cut}$ over 10 logarithmically spaced values between 100 and 100,000 keV. At each point of this grid, the spectral model is fitted independently and the smooth appearance of the plots results from the filled-contour visualization, which interpolates between the evaluated grid points.

\begin{figure}
    \centering
    \includegraphics[width=\linewidth]{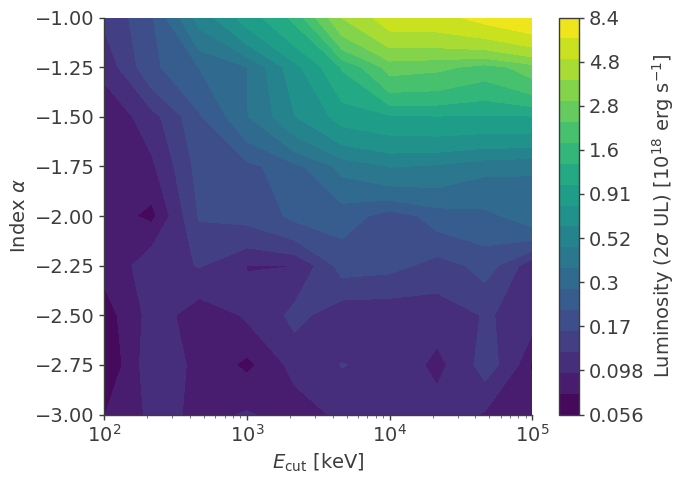}
    \caption{Upper limits on the luminosity obtained from the cutoff powerlaw (Eq.\,\ref{cutoffpowerlaw}).}
    \label{fig:Luminosity}
\end{figure}

\begin{figure}[h!]
    \centering
    \includegraphics[width=\linewidth]{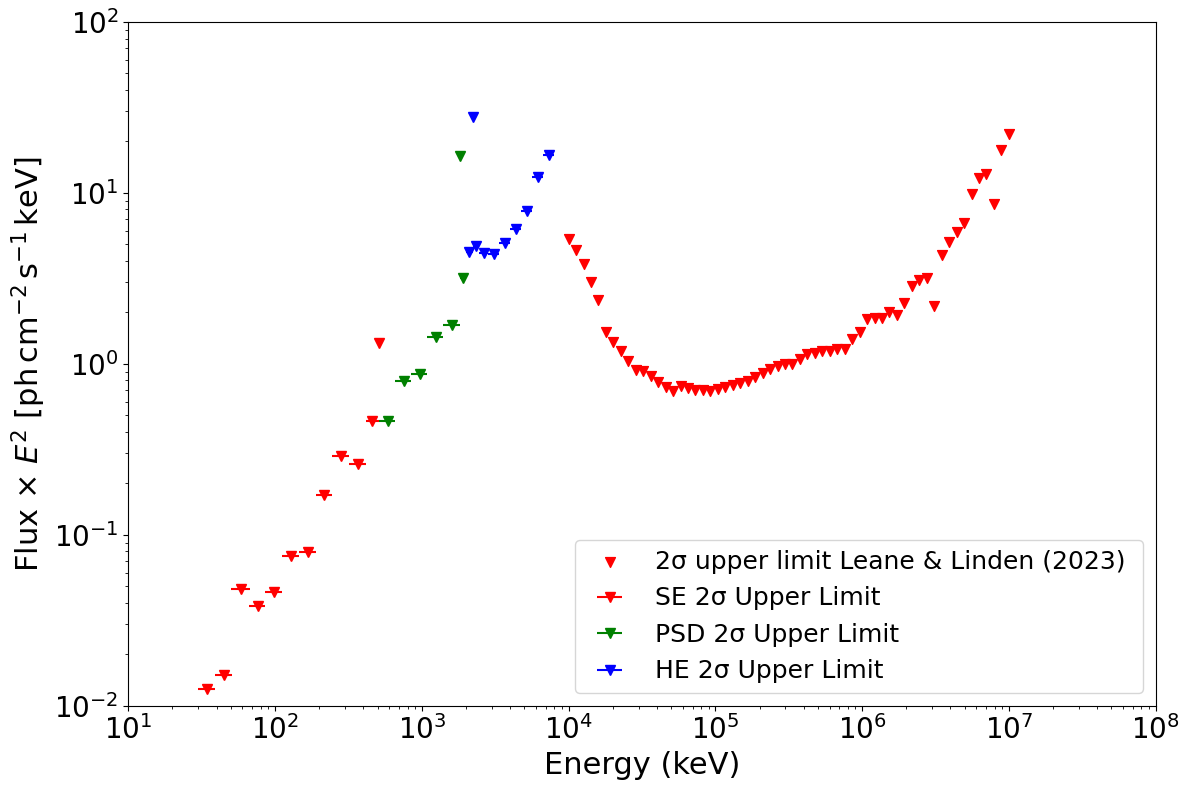}
    \caption{Upper limits for the data obtained with SPI, as well as, the data obtained through Fermi-LAT, taken from \citet{JupiterDM}.}
    \label{fig:SPI_plus_fermi}
\end{figure}

Fig.\,\ref{fig:Luminosity} shows the 2$\sigma$ calculated luminosity from the cutoff powerlaw model. Anything above this limit would have been detected by our instrument. We compare the data points from our measurements with the data from other instruments. In Fig.\,\ref{fig:SPI_plus_fermi} we show our upper limits together with data from Fermi-LAT taken from \citet{JupiterDM}.
The SPI upper limits appear to be compatible with the order of magnitude of the upper limits from the scanning instrument LAT. Around 10 MeV, the SPI and LAT measurements are compatible within a factor of a few.

\subsection{Constraints on reflected CGB, CR-induced emission, and dark matter}

\begin{figure}[h!]
        \centering
        \includegraphics[width=\linewidth]{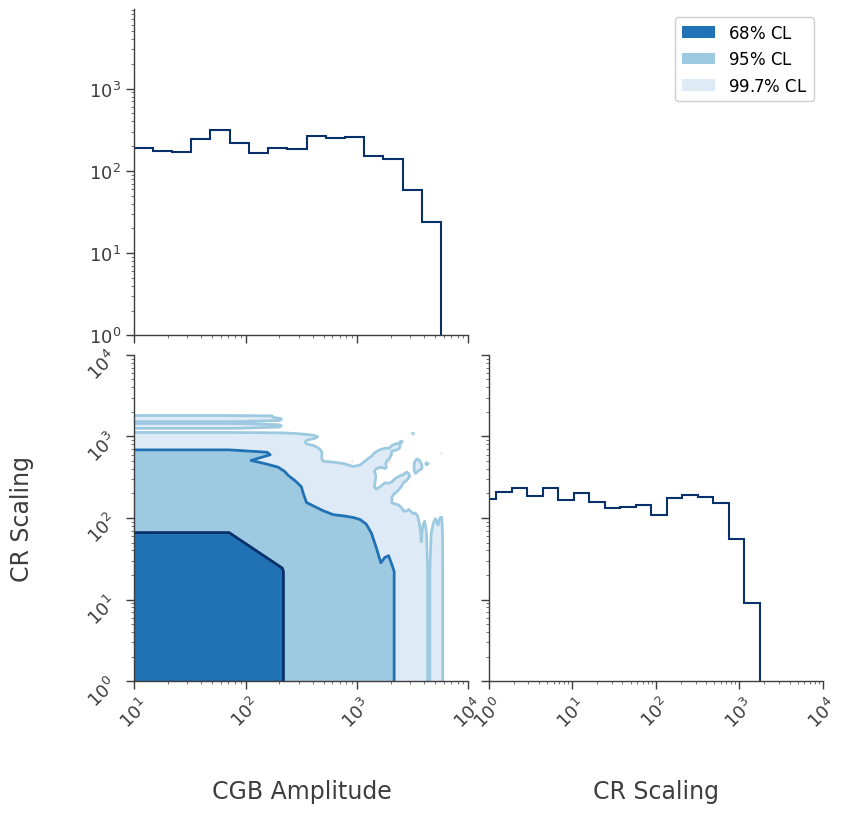}
        \caption{Joint posterior distribution of the CGB amplitude and the cosmic-ray scaling factor from the fit of the Jupiter atmospheric $\gamma$-ray spectrum. Off-diagonal panels show the 2D joint posterior, with shaded regions corresponding to the 68\%, 95\%, and 99.7\% credible regions. Diagonal panels show the 1D posteriors for each parameter.}
        \label{fig:Cornerplot}
\end{figure}

Fitting the amplitude of the CGB and the constant used to scale the total CR induced spectrum we obtain the constraints as shown in Fig.\,\ref{fig:Cornerplot}.

\begin{figure*}[!t]
    \centering
    \includegraphics[width=\linewidth]{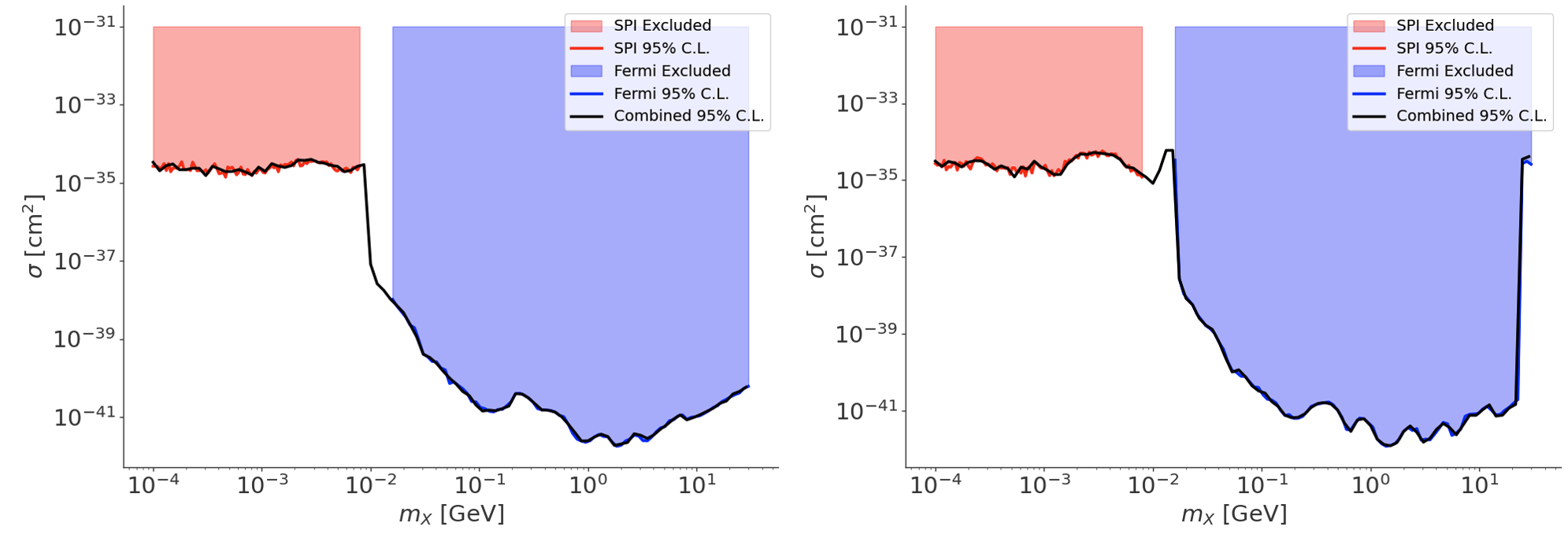}
    \caption{Exclusion region for dark matter–proton scattering cross section derived from INTEGRAL/SPI and Fermi-LAT observations of Jupiter, for the cases of $M_Y=0.1\times M_X$ (left) and $M_Y= M_X/1.01$ (right). Shaded regions signify the excluded regions by each instrument separately, while the black curve shows the combined 95\% confidence limit, from a combined treatment.}
    \label{fig:DMmodels}
\end{figure*}
The 2$\sigma$ upper limit for the CGB amplitude, obtained by fitting the reflected CGB model to the SPI data over 30 keV–8 MeV, is $9.2 \times 10^2$, while the value obtained from \citet{Ajello_2008} is $(10.15 \pm 0.80) \times 10^{-2}$, based on observations with the Burst Alert Telescope (BAT) aboard Swift in the 15–200 keV energy range. A detection of the CGB component from Jupiter would therefore require a signal approximately four orders of magnitude stronger than our current sensitivity. This indicates that our SPI analysis is not yet sensitive enough to detect a reflected CGB signal from Jupiter. The sensitivity of our measurements could be improved by the addition of previously excluded observations, due to their proximity to the Galactic center, given that we can model the background from other bright sources in the region appropriately. Furthermore, extending the analysis to additional Solar System bodies, such as Venus, Mars, or asteroid populations, may improve the overall sensitivity, allowing for better constraints in the CGB. Similarly, for the CR induced spectrum, the scaling factor has a 2$\sigma$ upper limit of $2.4 \times 10^2$. If the cosmic-ray-induced model were the correct description of the data, this factor would be expected to be unity. Therefore, our limit lies roughly two orders of magnitude above this, indicating that the present analysis is not yet sensitive to the predicted CR-induced flux level.

Regarding the constraints on the dark matter model, by fitting the model to the SPI data, as well as the data used in \citet{JupiterDM}, we were able to recreate the findings discussed in their work and extend the limits to the SPI region. This can be seen in Fig.\,\ref{fig:DMmodels}, where we provide results from the Fermi-LAT data as well our own. We further construct a unified exclusion framework spanning both the SPI and Fermi-LAT energy regimes, allowing for a continuous assessment of Jovian dark matter sensitivity. We test two cases for the mediator mass relative to the dark matter mass. In the first, the mediator is an order of magnitude lighter, $M_Y=0.1\times M_X$. In the second, it lies just below the dark matter mass, $M_Y = M_X/1.01$ The mediator mass sets the width of the energy box: from the cascade kinematics, a mediator mass close to $M_X$ produces a narrow box, while a lighter mediator produces a wide one. Because a narrow box deposits its flux into only a few energy bins, the resulting limit is set almost entirely by the data in those bins rather than the convolution over several energy bins. The two cases therefore affect the resulting exclusions region, as shown in Fig.\,\ref{fig:DMmodels}.


Thus, by combining INTEGRAL/SPI and Fermi-LAT observations, we demonstrate that Jupiter provides a continuous probe of long-lived mediator dark matter. However, while the combined SPI and Fermi-LAT exclusion extends to lower dark matter masses than previous Jovian studies, the precise low-mass sensitivity remains model dependent. In particular, dark matter evaporation inside Jupiter can substantially weaken the signal. Heavy-mediator scenarios generally predict evaporation thresholds in the $\mathrm{200MeV-1GeV}$ regime, while light-mediator models may allow sensitivity to persist to substantially lower masses \citep{JupiterDM}.
\section{Conclusion and Outlook} \label{Conclusions}

This work aimed to enhance the understanding of the CGB through the study of Jupiter's atmosphere as an intermediary to measure high-energy photons. Our aim was to measure the hard X-ray to soft $\gamma$-ray spectrum of Jupiter, using data acquired from the SPI instrument onboard the INTEGRAL satellite. In our methodology, we employed both the analysis of observational data from SPI and the simulation of photon and CR interactions with Jupiter's atmosphere. The analysis of SPI measurements involved several crucial steps to ensure the accuracy and reliability of the data. We cross correlated the SPI pointings for which Jupiter was coincidental in the SPI's field of view in the 22 years that INTEGRAL has been in operation. From these data, our model expectations were corrected for the relative distance between Earth and Jupiter. The final spectral data extracted from SPI represent, to our knowledge, the first measurement of a planetary atmosphere in the MeV range with a $\gamma$-ray telescope except Earth. All measured fluxes are consistent with zero, and we report 2$\sigma$ upper limits across the 30 keV to 8 MeV band. Fitting the data with a cutoff power law, a reflected-CGB component, a cosmic-ray-induced component, and a dark matter annihilation model, we derive 2$\sigma$ upper limits on the corresponding model amplitudes. The validity of the methodology is supported by the agreement of our fitted spectral parameters for the calibration source 3C\,273 ($\alpha = -1.64\pm0.13$,
$K = 1.7^{+1.2}_{-0.7}\times10^{-2}$) with literature values
\citep{Bouchet}.

The 2$\sigma$ upper limit we derive for the reflected CGB is four orders of magnitude above the expected given in literature,  if Jupiter reflects the spectrum given in \citet{Ajello_2008}. This can be attributed to the faintness of the reflected signal from Jupiter, due to the planet’s small angular size (30–50 arcseconds, depending on orbital configuration). The non-detection establishes a benchmark against which future measurements can be compared. Sensitivity could be further improved by including additional observations, specifically those near the Galactic Centre that were omitted in this work due to the strength of the sources
residing in that region.

The estimated intrinsic bremsstrahlung flux from Jupiter's aurora, $\sim 10^{-10}\,\mathrm{ph\,cm^{-2}\,s^{-1}\,keV^{-1}}$ at 20\,keV, lies five orders of magnitude below our upper limit at 30\,keV, confirming that intrinsic auroral emission is not a relevant foreground.

 Moreover, the scaling parameter from the CR induced spectrum is two orders of magnitude above unity. This shows that the measured signal is too faint to explain the induced spectrum, given the CR distribution around Jupiter taken from \citet{Boschini_2018} and \citet{Boschini_2020}. Here the sensitivity could also be improved by including more measurements, as the total exposure time of SPI towards Jupiter is 11.6\,MS whereas our dataset takes into account only 4.08\,Ms, suggesting an improvement of the sensitivity with SPI alone of $\sqrt{11.6/4.08}$. Something that is crucial to note is that the signal from the reflected CGB and the one from the CR induced emission, differ distinctly from each other. While the reflected component scales purely with the inverse-square distance between Earth and Jupiter, the cosmic-ray-induced component is additionally modulated by the 11-year solar cycle, since the cosmic-ray flux impinging on Jupiter's atmosphere depends on the heliospheric modulation potential.

The dark matter exclusion region obtained demonstrates the ability to utilise Jupiter as a probe of the long-lived mediator scenario over a continuous mass range. Our SPI-derived constraint extends the Jovian dark matter exclusion of \citet{JupiterDM} to lower dark matter masses, in the regime where the resulting box-shaped spectrum from $\chi\bar\chi \to \phi\phi \to 4\gamma$ falls within the SPI energy range. For both the cases of a light mediator ($M_Y = 0.1\times M_X$), and for the near-degenerate case ($M_Y = M_X/1.01$), SPI excludes cross sections with the best constraint at $\sigma \sim [10^{-35}] \,\mathrm{cm}^2$ over the mass range $m_X \sim [10^{-2}]$--$[10^{-4}]\,\mathrm{GeV}$.

Future MeV missions will substantially improve this situation. In particular, the Compton Spectrometer and Imager (COSI) \citep{Tomsick2025}, scheduled for launch in 2027, will repeatedly survey the sky along the ecliptic, ensuring that planets, the Moon, and a large fraction of the asteroid population remain frequently within the field of view. Effective exposure times along the ecliptic are expected to reach 31–62Ms, substantially exceeding the exposure available in the present SPI analysis. Combined with COSI’s improved continuum sensitivity near 511\,keV, where pair production in planetary atmospheres produces positron-annihilation radiation, this could reduce statistical uncertainties by at least an order of magnitude and potentially enable the first detection of Solar System $\gamma$-ray albedo signals.

The moving-source analysis pipeline, atmospheric response framework, and forward-folding machinery developed in this work is directly applicable to other Solar System bodies. Venus, owing to its proximity to Earth and angular diameter of up to 60 arcseconds, represents a promising next target for archival INTEGRAL analyses. Beyond individual planets, the collective contribution of asteroid populations and potentially the Oort cloud \citep{moskalenko2009} may provide a diffuse Solar System scattering component, effectively increasing the total solid angle available for reflecting the CGB. This will offer more data and allow for tighter constraints on the diffuse MeV background.

We have demonstrated the viability of using Jupiter as a probe of the diffuse cosmic $\gamma$-ray background in the hard X-ray to soft MeV energy range. While the present upper limits do not constrain the CGB or cosmic-ray-induced components individually, the methodology developed is robust, as shown by the validation with known sources, and it is directly applicable to other Solar System bodies and future instruments. With improved background modelling near the Galactic centre and anticentre, the inclusion of additional targets, and the increased sensitivity of forthcoming MeV $\gamma$-ray telescopes, planetary reflection techniques offer a clear path toward the first measurement of the MeV CGB via planetary reflectors.


\bibliographystyle{aa}
\bibliography{refrences}

\begin{acknowledgements}
Dimitris Tsatsis and Tristan Bouchet acknowledge support from the
\emph{DFG/LIS} project SI\,2502/6-1, project number 551127478.
Rudi Reinhardt acknowledges support by the
\emph{Bundesministerium für Forschung, Technologie und Raumfahrt}
as part of the
\emph{FORKA (Forschung für den Rückbau kerntechnischer Anlagen)}
funding program under reference number 15S9455\,A--E.
Saurabh Mittal acknowledges support by the
\emph{Bundesministerium für Wirtschaft und Energie via the Deutsches
Zentrum für Luft- und Raumfahrt (DLR)}
under Contract No.~50\,OO\,2219.
Laura Eisenberger acknowledges support by the
\emph{Bundesministerium für Wirtschaft und Energie via the Deutsches
Zentrum für Luft- und Raumfahrt (DLR)}
under Contract No.~50\,OR\,2413 and is grateful for the support of the
\emph{Studienstiftung des Deutschen Volkes}.
Tomohiko Oka acknowledges support by the
\emph{JSPS Overseas Research Fellowship}.\

We thank Rebecca K.~Leane and Tim Linden for kindly providing the data
used in this work from their study of gamma-ray emission from Jupiter \citep{JupiterDM}.
\end{acknowledgements}

\onecolumn
\appendix
\section{CR Simulation Results} \label{AppendixCR}
\begin{figure*}[!h] 
    \centering
    \begin{subfigure}[b]{0.47\textwidth}
        \centering
        \includegraphics[
            width=\textwidth]{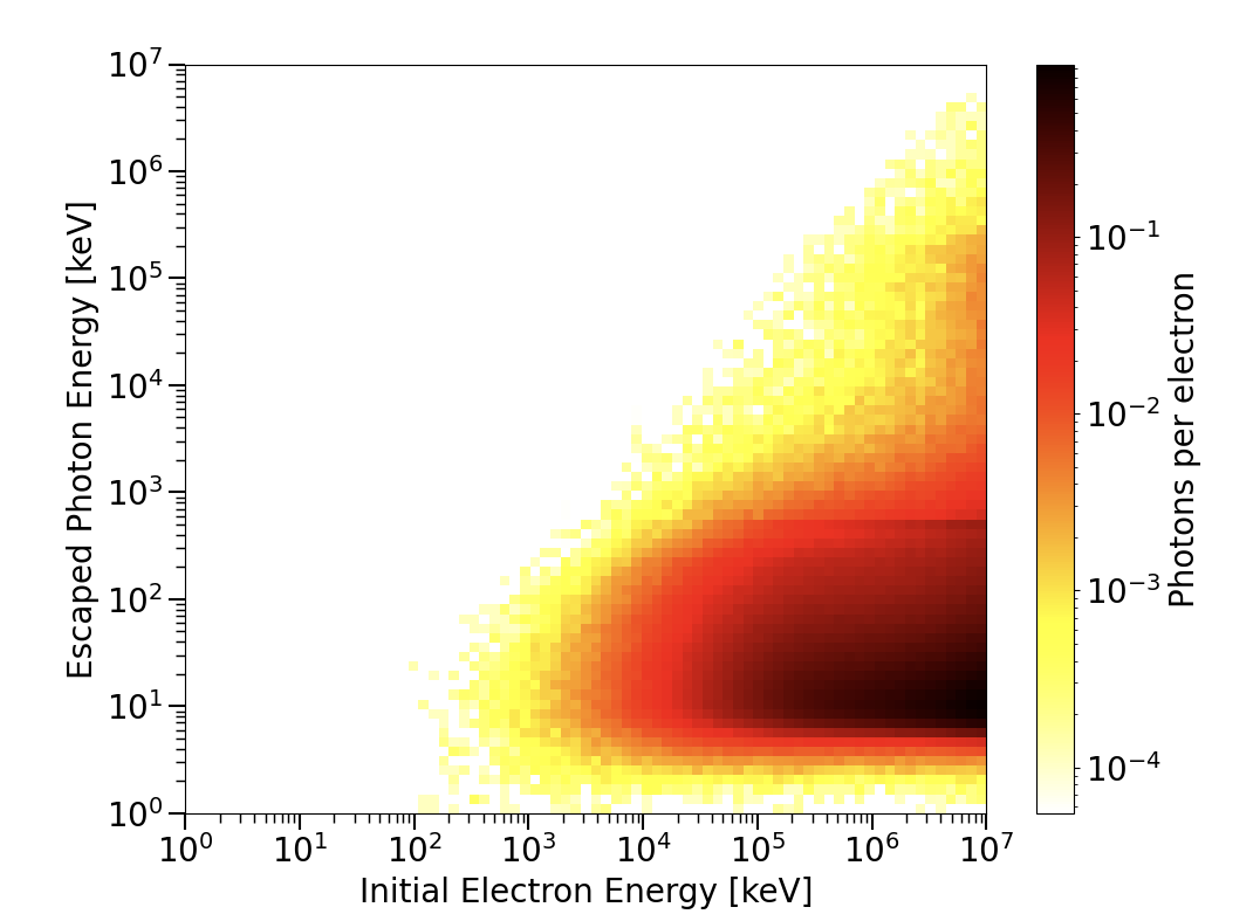}
        \caption{Electron energy dispersion matrix.}
        \label{fig:edisp_e}
    \end{subfigure}
    \hfill
    \begin{subfigure}[b]{0.47\textwidth}
        \centering
        \includegraphics[width=\textwidth]{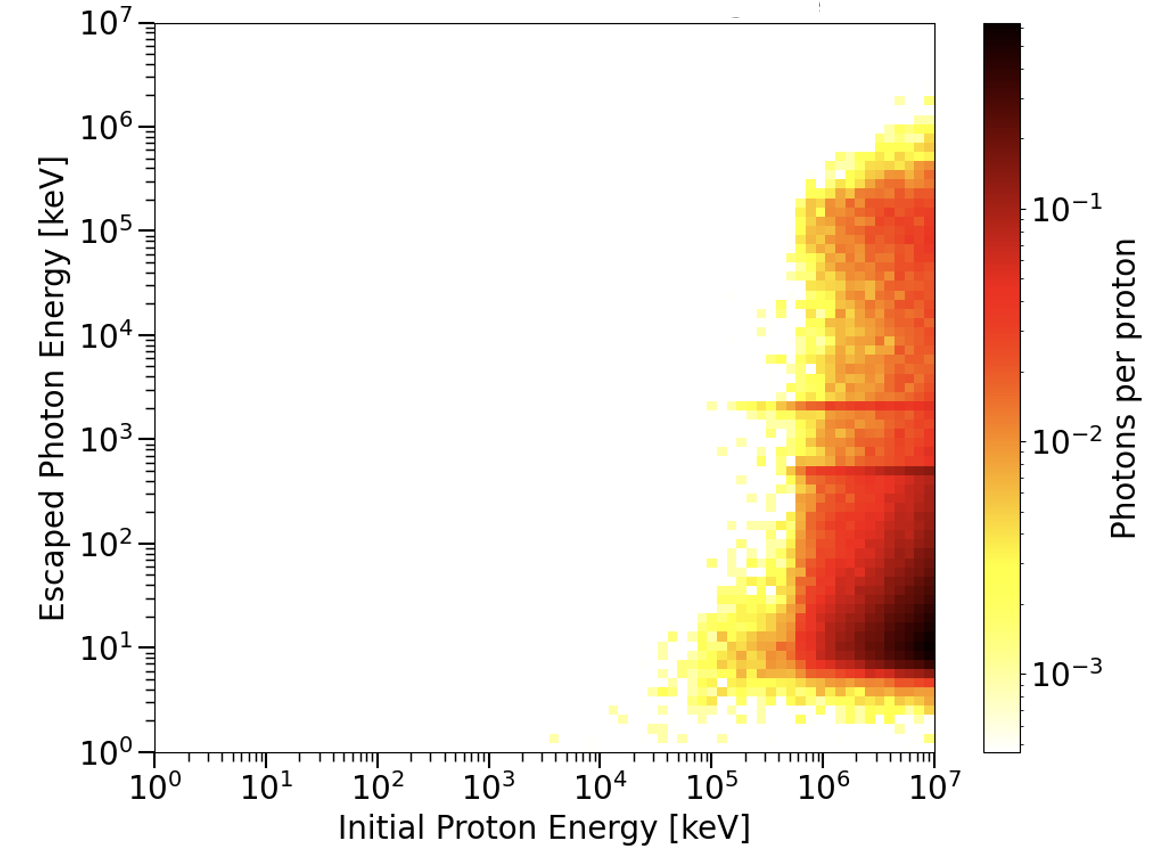}
        \caption{Proton energy dispersion matrix.}
        \label{fig:edisp_p}
    \end{subfigure}

    \vspace{0.5cm}

    \begin{subfigure}[b]{0.47\textwidth}
        \centering
        \includegraphics[width=\textwidth]{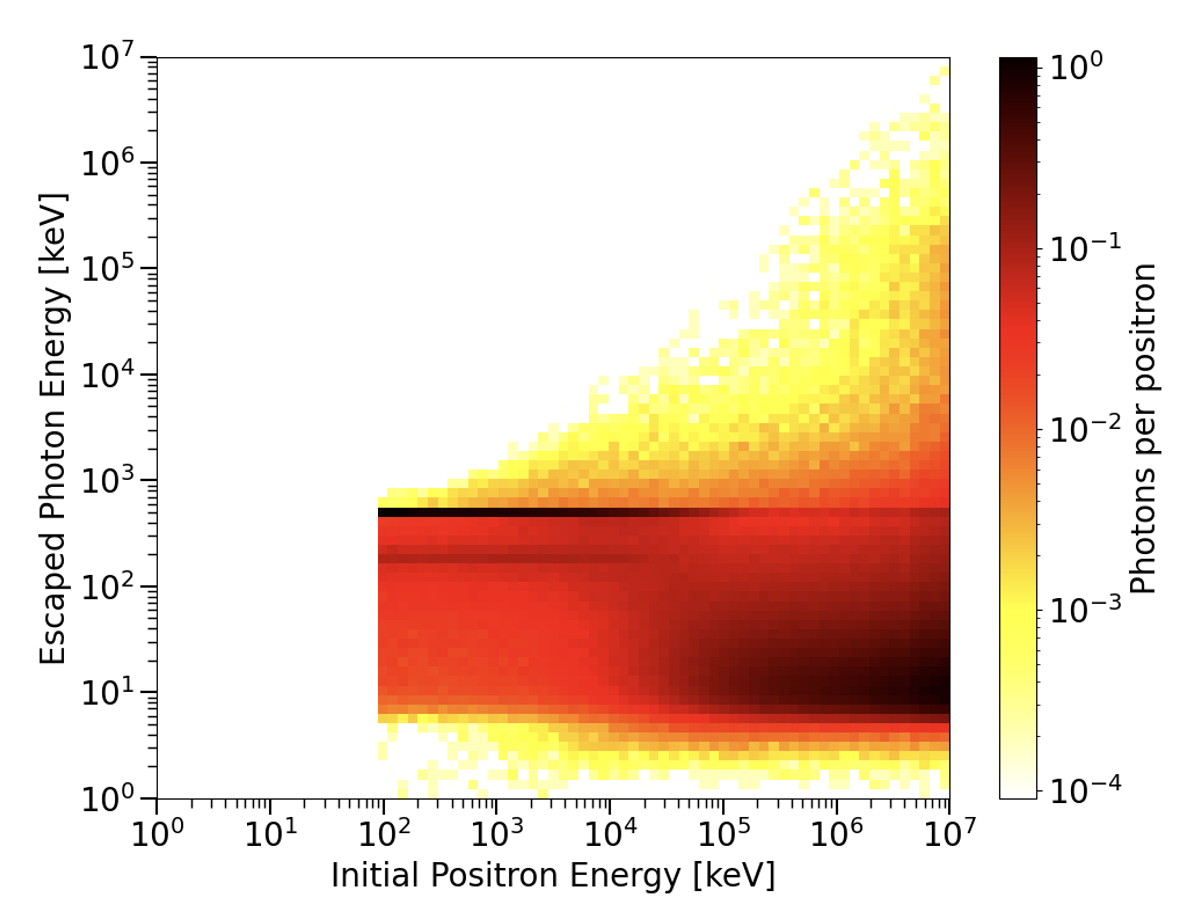}
        \caption{Positron energy dispersion matrix.}
        \label{fig:edisp_a}
    \end{subfigure}
    \hfill
    \begin{subfigure}[b]{0.47\textwidth}
        \centering
        \includegraphics[width=\textwidth]{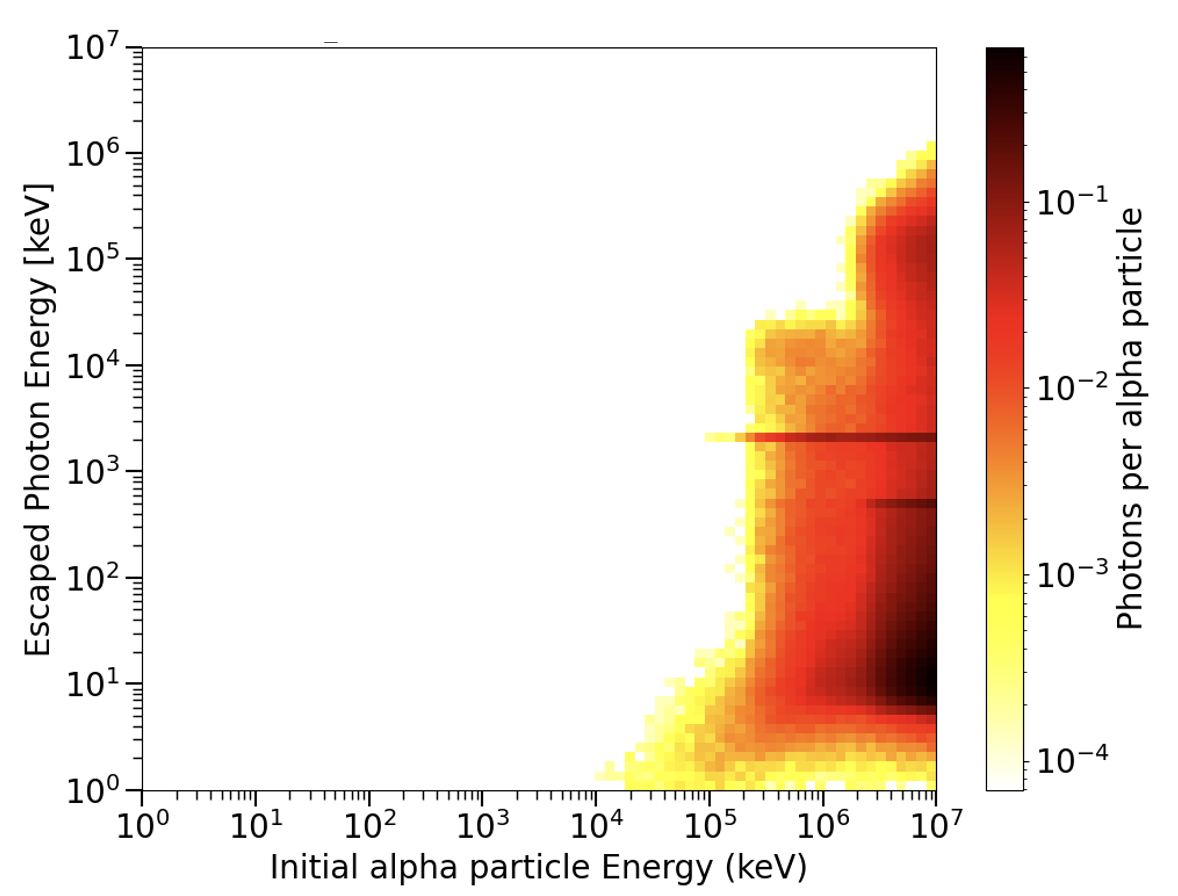}
        \caption{Alpha particles energy dispersion matrix.}
        \label{fig:edisp_pos}
    \end{subfigure}

    \caption{Atmospheric response matrices for the four cosmic-ray species. 
             Each matrix shows the probability of an incident particle of 
             energy $E_\mathrm{in}$ producing a photon of energy $E_\gamma$.}
    \label{fig:edisp_all}
\end{figure*}

In Fig.\,\ref{fig:edisp_all} we show the atmospheric yield for the interaction of CR (electrons, protons, positron, alpha) with Jupiter's atmosphere.

\twocolumn
\end{document}